\documentclass[prb,aps,nofootinbib,nobibnotes,notitlepage,superscriptaddress,twocolumn]{revtex4-1}

\usepackage{amsthm}
\usepackage{amsmath,bm}
\usepackage{amssymb}
\usepackage{amsfonts}
\usepackage{graphicx}

\usepackage{txfonts}

\usepackage{float}

\usepackage{bbm}
\newcommand{\expect}[1]{\mathinner{\langle #1\rangle}}
\newcommand\abs[1]{\lvert#1\rvert}
\DeclareMathOperator{\tr}{Tr}
\newcommand{\identity}{\mathbbm{1}}
\newcommand{\id}{\identity}
\newcommand{\ox}{\otimes}
\newcommand{\bra}[2][]{%
  \def\@tempa{#1}%
  \ifx\@tempa\@empty\relax%
    \mathinner{\langle #2\rvert}%
  \else%
    \mathinner{\langle #2\rvert}_{#1}%
  \fi}
\newcommand{\ket}[2][]{%
  \def\@tempa{#1}%
  \ifx\@tempa\@empty\relax%
    \mathinner{\lvert#2\rangle}%
  \else%
    \mathinner{\lvert#2\rangle}_{\hspace{-0.1em}#1}%
  \fi}
\newcommand{\ketbra}[3][]{%
  \def\@tempa{#1}%
  \ifx\@tempa\@empty\relax%
    \mathinner{\lvert#2\rangle\langle #3\rvert}%
  \else%
    \mathinner{\lvert#2\rangle\langle #3\rvert}_{#1}%
  \fi}
\newcommand{\proj}[2][]{\ketbra[#1]{#2}{#2}}
\newcommand{\dd}{\mathrm{d}}
\newcommand\braXket[4][]{\mathinner{\langle#2\vert#3\vert#4\rangle}_{#1}}
\newcommand{\braket}[2]{{\langle}{#1}|{#2}{\rangle}}

\usepackage[colorlinks=true,linkcolor=blue,citecolor=blue,urlcolor=blue]{hyperref}

\newtheorem{lem}{Lemma}
\newtheorem{thm}{Theorem}
\newtheorem{cor}{Corollary}

\begin{document}

\newcommand{\avg}[1]{\langle#1\rangle}		
\newcommand{\var}{\mathrm{Var}}		

\newcommand{\Vs}[1]{\var[#1]}
\newcommand{\cvar}{\var_{\mathrm{Q}}^{\mathrm{B}|\mathrm{A}}}

\newcommand{\ie}{\textit{i.e.\ }}

\renewcommand{\vec}[1]{\mathbf{#1}}
\newcommand{\bg}[1]{\boldsymbol{#1}}
\newcommand{\inlineheading}[1]{\textit{{#1.---}}}
\newcommand{\mc}[1]{\mathcal{#1}}
\newcommand{\mb}[1]{\mathbf{#1}}
\newcommand{\comment}[1]{{\color{red}[#1]}}
\newcommand{\fisher}{F}
\newcommand{\cfi}{\fisher}
\newcommand{\qfi}{\fisher_{\mathrm{Q}}}
\newcommand{\cqfi}{\fisher_{\mathrm{Q}}^{\mathrm{B}|\mathrm{A}}}

\newcommand{\assem}{\mc{A}}
\newcommand{\tqfi}{\bar{\mc{F}}}
\newcommand{\tcqfi}{\bar{\mc{F}}^{\mathrm{B}|\mathrm{A}}}
\newcommand{\steer}{\mc{S}}
\newcommand{\steerAvg}{\steer_\mathrm{avg}}
\newcommand{\steerMax}{\steer_\mathrm{max}}
\newcommand{\pos}[1]{\left[ #1 \right]^+}

\newcommand{\mg}[1]{{\color{cyan}[(MG) #1]}}
\newcommand{\by}[1]{{\color{red}[(BY) #1]}}
\newcommand{\mf}[1]{{\color{blue}[(MF) #1]}}

\title{Metrological complementarity reveals the Einstein-Podolsky-Rosen paradox}

\author{Benjamin Yadin}
\email{benjamin.yadin@gmail.com}
    \affiliation{School of Mathematical Sciences and Centre for the Mathematics and Theoretical Physics of Quantum Non-Equilibrium Systems,
University of Nottingham, University Park, Nottingham NG7 2RD, United Kingdom}
    \affiliation{Wolfson College, University of Oxford, Linton Road, Oxford OX2 6UD, United Kingdom}
\author{Matteo Fadel}
\email{matteo.fadel@unibas.ch}
\affiliation{Department of Physics, University of Basel, Klingelbergstrasse 82, 4056 Basel, Switzerland}
\author{Manuel Gessner}
\email{manuel.gessner@ens.fr}
\affiliation{Laboratoire Kastler Brossel, ENS-Universit\'{e} PSL, CNRS, Sorbonne Universit\'{e}, Coll\`{e}ge de France, 24 Rue Lhomond, 75005, Paris, France}

\date{\today}

\maketitle

\section*{Abstract}

\textbf{
The Einstein-Podolsky-Rosen (EPR) paradox plays a fundamental role in our understanding of quantum mechanics, and is associated with the possibility of predicting the results of non-commuting measurements with a precision that seems to violate the uncertainty principle. This apparent contradiction to  complementarity is made possible by nonclassical correlations stronger than entanglement, called steering. 
Quantum information recognises steering as an essential resource for a number of tasks but, contrary to entanglement, its role for metrology has so far remained unclear. Here, we formulate the EPR paradox in the framework of quantum metrology, showing that it enables the precise estimation of a local phase shift and of its generating observable. Employing a stricter formulation of quantum complementarity, we derive a criterion based on the quantum Fisher information that detects steering in a larger class of states than well-known uncertainty-based criteria. Our result identifies useful steering for quantum-enhanced precision measurements and allows one to uncover steering of non-Gaussian states in state-of-the-art experiments.
}

\section*{Introduction}

In their seminal 1935 paper~\cite{EPR1935}, EPR presented a scenario where the position and momentum of one quantum system (B) can both be predicted with certainty from local measurements of another remote system (A). Based on this apparent violation of the uncertainty principle, in 1989 Reid formulated the first practical criterion for an EPR paradox~\cite{ReidPRA1989}, which has enabled numerous experimental observations~\cite{ReidRMP2009}: Steering from A to B is revealed when measurement results of A allow to predict the measurement results of B with errors that are smaller than the limit imposed by the Heisenberg-Robertson uncertainty relation for B. More generally, an EPR paradox implies the failure of any attempt to describe the correlations between the two systems in terms of classical probability distributions and local quantum states for B, known as local hidden state (LHS) models, as was shown by Wiseman et al.\ in 2007 using the framework of quantum information theory~\cite{WisemanPRL2007}. Aside from its fundamental interest, steering is recognised as an essential resource for quantum information tasks~\cite{GuehneRMP2020}, such as one-sided device-independent quantum key distribution~\cite{BranciardPRA2012,AolitaPRX2015} and quantum channel discrimination~\cite{PianiPRL2015}. 

Uncertainty relations describe the complementarity of non-commuting observables, but the complementarity principle applies more generally to notions that are not necessarily associated with an operator. One generalisation~\cite{BraunsteinAPhys1996} involves the quantum Fisher information (QFI), the central tool for quantifying the precision of quantum parameter estimation~\cite{Paris2009,GiovannettiNatPhoton2011,TothJPA2014,PezzeRMP2018}. Besides its fundamental relevance for quantum-enhanced precision measurements, the QFI is of great interest for the characterisation of quantum many-body systems~\cite{HaukeNatPhys2016,PezzePRL2017} and gives rise to an efficient and experimentally accessible witness for multipartite entanglement~\cite{TothJPA2014,PezzeRMP2018,StrobelSCIENCE2014}, but so far, its relation to steering has remained elusive. It has been a long-standing open problem to determine if quantum correlations stronger than entanglement, such as steering or Bell correlations, play a role in metrology~\cite{Frowis19}.

In this work we formulate a steering condition in terms of the complementarity of a phase shift $\theta$ and its generating Hamiltonian $H$, using information-theoretic tools from quantum metrology. We express our steering condition in terms of the QFI. The more general phase-generator complementarity principle reproduces the Heisenberg-Robertson uncertainty relation in the special case where the phase is estimated from an observable $M$. Therefore, our metrological criterion is stronger than the uncertainty-based approach and allows us to uncover hidden EPR paradoxes in experimentally relevant scenarios. Our result answers positively the question of whether steering can be a resource in quantum sensing applications.
\\

\begin{figure}[tb]
\centering
\includegraphics[width=.49\textwidth]{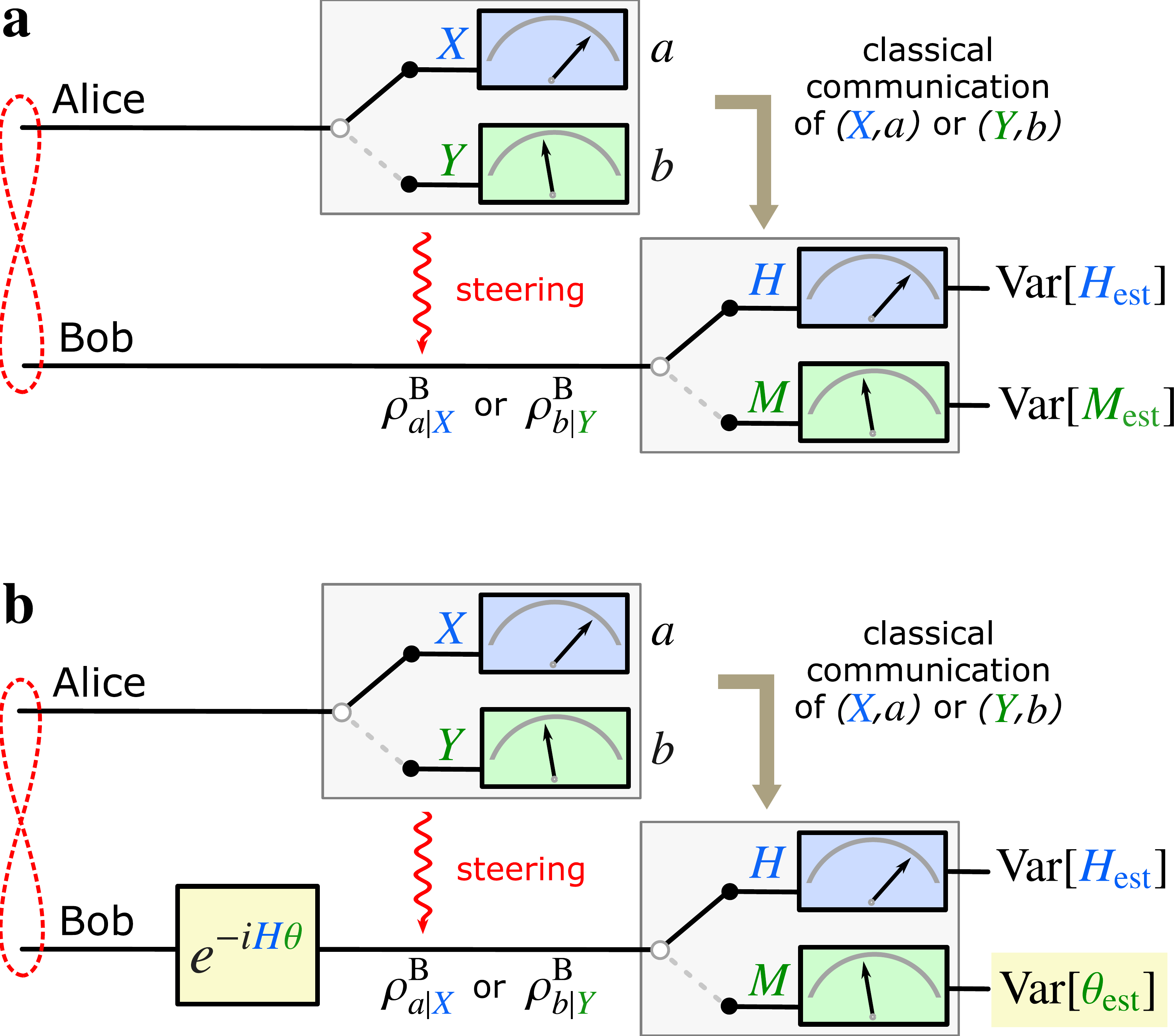}
\caption{\textbf{Formulation of the EPR paradox as a metrological task.} \textbf{a)} In the standard EPR scenario, Alice's measurement setting $X$ ($Y$), and result $a$ ($b$), leave Bob in the conditional quantum states $\rho^{\mathrm{B}}_{a|X}$ ($\rho^{\mathrm{B}}_{b|Y}$). Knowing Alice's setting and result allows Bob to choose what
measurement to perform on his state, and to make a prediction for the result. In an ideal scenario with strong quantum correlations, Alice's measurement of $X$ ($Y$) steers Bob into an eigenstate of his observable $H$ ($M$), allowing him to predict the result with certainty. When $H$ and $M$ do not commute, this seems to contradict the complementarity principle. In practice, an EPR paradox is revealed whenever Bob's predictions are precise enough to observe an apparent violation of Heisenberg's uncertainty relation, see Eq.~(\ref{eq:Reid}).
\textbf{b)} In our formulation of the EPR paradox as a metrological task, a local phase shift $\theta$ is generated by $H$ on Bob's state. Then, depending on Alice's measurement setting and result, he decides whether to predict and measure $H$ (as before), or to estimate $\theta$ from the measurement $M$. Here, Bob can choose the observable $M$ as a function of Alice's measurement result. The complementarity between $\theta$ and its generator $H$ seems to be contradicted if the lower bound on their estimation errors, Eq.~(\ref{eq:steeringcondphasegen}), is violated. This gives a metrological criterion for observing the EPR paradox. Since the metrological complementarity is sharper than the uncertainty-based notion, this approach leads to a tighter criterion to detect steering. Both results coincide in the special case when Bob estimates $\theta$ only from the observable $M$.}
\label{fig:1}
\end{figure}

\section*{Results}

\textbf{Reid's criterion for an EPR paradox.}
We first recall some basic definitions by considering the following scenario (see Fig.~\ref{fig:1}\textbf{a}). Alice (A) performs on her subsystem a measurement and communicates her setting $X$ and result $a$ to Bob (B). Based on this information, Bob uses an estimator $h_{\mathrm{est}}(a)$ to predict the result of his subsequent measurement of $H=\sum_hh|h\rangle\langle h|$. The average deviation between the prediction and Bob's actual result $h$ is given by
$	\Vs{H_\mathrm{est}}  := \sum_{a,h} p(a,h|X,H) \left( h_{\mathrm{est}}(a) - h \right)^2$,
often called the inference variance~\cite{ReidRMP2009},
where $p(a,h|X,H)$ is the joint probability distribution for results $a$ and $h$, conditioned on the measurement settings $X$ and $H$. The procedure is repeated with different measurement settings $Y$ and $M$, and Reid's criterion~\cite{ReidPRA1989,ReidRMP2009} for an EPR paradox consists of a violation of the local uncertainty limit
\begin{align}\label{eq:Reid}
\Vs{H_\mathrm{est}}\Vs{M_\mathrm{est}}\geq\frac{|\langle [H,M]\rangle_{\rho^{\mathrm{B}}}|^2}{4}.
\end{align}
From the perspective of quantum information theory, the condition~(\ref{eq:Reid}) plays the role of a witness for steering, but it may not always succeed in revealing an EPR paradox.

The most general way to formally model the joint statistics $p(a,h|X,H)$ is offered by the formalism of assemblages, \ie functions $\assem(a,X) = p(a|X) \rho^{\mathrm{B}}_{a|X}$ that map any possible result $a$ of Alice's measurement of $X$ to a local probability distribution $p(a|X)$ and a (normalised) conditional quantum state $\rho^{\mathrm{B}}_{a|X}$ for Bob's subsystem~\cite{CavalcantiREPPROGPHYS2017}. This description avoids the need to make assumptions about the nature of Alice's system, which is key to one-sided device-independent quantum information processing~\cite{BranciardPRA2012,AolitaPRX2015,GuehneRMP2020}. We only impose a no-signalling condition which requires that $\sum_a \assem(a,X) = \rho^{\mathrm{B}}$ for all $X$, where $\rho^{\mathrm{B}}$ is the reduced density matrix of Bob's system. Based on the assemblage $\mathcal{A}$, the joint statistics are described as $p(a,h|X,H)=p(a|X)\langle h|\rho^{\mathrm{B}}_{a|X}|h\rangle$.

The EPR paradox can now be formally defined as an observation that rules out the possibility of modelling an assemblage by a local hidden state (LHS) model. In such a model, a classical random variable $\lambda$ with probability distribution $p(\lambda)$ determines both Alice's statistics $p(a|X,\lambda)$ and Bob's local state $\sigma^{\mathrm{B}}_\lambda$, leading to the assemblage $\assem(a,X) = \sum_\lambda p(a|X,\lambda) p(\lambda) \sigma^{\mathrm{B}}_\lambda$. Inequality~(\ref{eq:Reid}) holds for arbitrary estimators and measurement settings whenever a LHS model exists. The sharpest formulation of Eq.~(\ref{eq:Reid}) is thus obtained by optimising these choices to minimise the estimation error. The optimal estimator $h_{\mathrm{est}}(a)=\mathrm{Tr}\{\rho^{\mathrm{B}}_{a|X}H\}$ attains the lower bound~\cite{ReidRMP2009}  $\Vs{H_\mathrm{est}} \geq  \sum_{a} p(a|X) \var[\rho^{\mathrm{B}}_{a|X},H]$, where $\var[\rho,H]=\langle H^2\rangle_{\rho}-\langle H\rangle_{\rho}^2$ is the variance with $\langle O\rangle_{\rho}=\mathrm{Tr}\{\rho O\}$. Optimising over Alice's measurement setting $X$ leads to the quantum conditional variance
\begin{equation}\label{eq:optcondvar}
	\cvar[\assem,H] := \min_X \, \sum_a p(a|X) \var[\rho^{\mathrm{B}}_{a|X},H],
\end{equation}
and the optimised version of Reid's condition~(\ref{eq:Reid}) reads
$\cvar[\assem,H]\cvar[\assem,M]\geq |\langle [H,M]\rangle_{\rho^{\mathrm{B}}}|^2/4$.
The uncertainty-based detection of the EPR paradox is based on the fact that Alice's choice of measurement can steer Bob's system into conditional states that have small variances for either one of the two non-commuting observables $H$ and $M$.\\

\textbf{EPR-assisted metrology.}
To express quantum mechanical complementarity in the framework of quantum metrology~\cite{Paris2009,GiovannettiNatPhoton2011,TothJPA2014,PezzeRMP2018}, we assume that the observable $H$ imprints a local phase shift $\theta$ on Bob's system through the unitary evolution $e^{-iH\theta}$ -- see Fig.~\ref{fig:1}\textbf{b}. The phase shift $\theta$ is complementary to the generating observable $H$ and we show that the violation of
\begin{align}\label{eq:steeringcondphasegen}
\Vs{\theta_{\mathrm{est}}}\Vs{H_{\rm{est}}}\geq \frac{1}{4n}
\end{align}
implies an EPR paradox and reveals steering from $A$ to $B$. Here, $\Vs{\theta_{\mathrm{est}}}$ describes the error of an arbitrary estimator for the phase $\theta$, constructed from local measurements by Alice and Bob on $n$ copies of their state. Given any $M$, it is possible to construct an estimator $\theta_{\mathrm{est}}$ that achieves in the central limit ($n\gg 1$)
\begin{align}\label{eq:variancebound}
\Vs{\theta_{\mathrm{est}}} = \frac{\Vs{M_{\rm{est}}}}{n|\langle [H,M]\rangle_{\rho^{\mathrm{B}}}|^2}.
\end{align}
Essentially, we convert an $n$-sample average of $M_\mathrm{est}$ into an estimate of $\theta$ (see Methods for details).
For this specific estimation strategy, we thus recover the uncertainty-based formulation~(\ref{eq:Reid}) of the EPR paradox from the more general expression~(\ref{eq:steeringcondphasegen}).

In the following, we will derive our main result, which will allow us to prove the above statements. First note that the local phase shift acts on Bob's conditional quantum states but has no impact on Alice's measurement statistics due to no-signalling, and thus produces the assemblage $\assem_{\theta}(a,X) = p(a|X) \rho^{\mathrm{B}}_{a|X,\theta}$, where $\rho^{\mathrm{B}}_{a|X,\theta}=e^{-iH\theta}\rho^{\mathrm{B}}_{a|X}e^{iH\theta}$. This implies the phase shift has no impact on the existence of LHS models, and $\assem_{\theta}(a,X) = \sum_\lambda p(a|X,\lambda) p(\lambda) \sigma^{\mathrm{B}}_{\lambda,\theta}$. Without any assistance from Alice, Bob's precision of the estimation of $\theta$ is determined by his reduced density matrix $\rho^{\mathrm{B}}_\theta$. In this case, the error of an arbitrary unbiased estimator $\theta_{\mathrm{est}}^{\mathrm{B}}$ for $\theta$ is bounded by the quantum Cram\'er-Rao bound, $\Vs{\theta^{\mathrm{B}}_{\mathrm{est}}}\geq (n\qfi[\rho^{\mathrm{B}},H])^{-1}$, the central theorem of quantum metrology~\cite{Holevo,Helstrom,BraunsteinPRL1994,GiovannettiNatPhoton2011,TothJPA2014,PezzeRMP2018}, where $\qfi[\rho^{\mathrm{B}},H]$ is the QFI. $\qfi[\rho^{\mathrm{B}},H]$ can be thought of intuitively as measuring the sensitivity of the state $\rho^{\mathrm{B}}$ to evolution generated by $H$; see Methods for a formal definition and explicit expression in terms of the eigenvectors and eigenvalues of $\rho^{\mathrm{B}}$. The quantum Cram\'er-Rao bound can be saturated by optimising both the estimator and the measurement observable~\cite{BraunsteinPRL1994}.

In the assisted phase-estimation protocol, Fig.~\ref{fig:1}\textbf{b}, Alice communicates to Bob her measurement setting and result, \ie $X$ and $a$. This additional knowledge allows Bob to adapt the choice of his observable as a function of the conditional state $\rho^{\mathrm{B}}_{a|X}$ and to achieve the maximal sensitivity $F_{\mathrm{Q}}[\rho^{\mathrm{B}}_{a|X},H]$ for an estimation of $\theta$. This way, he can attain an average sensitivity as large as the quantum conditional Fisher information
\begin{align} \label{eq:assisted_qfi}
	\cqfi[\assem,H] 	& := \max_X \, \sum_a p(a|X) \qfi[\rho^{\mathrm{B}}_{a|X},H].
\end{align}
As the main result of our paper, we show that in the absence of steering the quantum conditional Fisher information~(\ref{eq:assisted_qfi}) is always bounded from above in terms of the quantum conditional variance~(\ref{eq:optcondvar}): For any assemblage $\assem$ that admits a LHS model, the following bound holds:
\begin{align} \label{eq:main}
	\cqfi[\assem,H] \leq 4 \cvar[\assem,H].
\end{align}
The proof (see Methods) primarily follows from the fact that the QFI $\qfi[\rho,H]$ is a convex function of the state $\rho$, while the variance $\var[\rho,H]$ is instead concave~\cite{TothPRA2013}.
Note that $\qfi[\rho^{\mathrm{B}},H]\leq 4\var[\rho^{\mathrm{B}},H]$ holds for arbitrary $\rho^{\mathrm{B}}$, and by means of the Cram\'er-Rao bound implies the phase-generator complementarity relation
\begin{align}
    \var[\theta_{\mathrm{est}}^{\mathrm{B}}]\var[\rho^{\mathrm{B}},H]\geq \frac{1}{4n}.
\end{align}
This clearly shows how a violation of~(\ref{eq:steeringcondphasegen}) implies an EPR paradox. The result~(\ref{eq:main}) has several important consequences that we discuss in the remainder of this article.

Useful steering for quantum metrology is identified by correlations that violate the condition~(\ref{eq:main}). We note that classical correlations between Alice and Bob may be sufficient for having $\qfi[\rho^{\mathrm{B}},H] < \cqfi[\rho^{\mathrm{AB}},H]$ and $\cvar[\rho^{\mathrm{AB}},H] < \var[\rho^{\mathrm{B}},H]$. This shows that assistance is useful even in the absence of steering to improve the estimation precision for $\theta$ and $H$, but only with steering can the limit defined by quantum mechanical complementarity~(\ref{eq:main}) be overcome.\\

\textbf{Comparison to Reid-type criteria.}
The metrological steering condition~(\ref{eq:main}) is stronger than standard criteria based on Heisenberg-Robertson uncertainty relations. In fact, the lower bound
\begin{align}\label{eq:lbound}
 \frac{ \abs{\expect{[H,M]}_{\rho^{\mathrm{B}}}}^2 }{\cvar[\assem,M]}\leq \cqfi[\assem,H]
 \end{align}
holds for arbitrary observables $H,M$ and, besides no-signalling, does not require assumptions about the assemblage $\assem$ (see Methods for the proof). Hence, the bound~(\ref{eq:main}) implies Reid's uncertainty-based condition~(\ref{eq:Reid}) for all LHS models. In experimentally relevant situations where the observables $H$ and $M$ are chosen as linear observables, such as quadrature measurements in quantum optics or collective spins in atomic systems, the bound~(\ref{eq:lbound}) can be interpreted as a Gaussian approximation to the assisted sensitivity. In fact, violation of criterion~\eqref{eq:Reid} (choosing the appropriate observables) is necessary and sufficient for steering of Gaussian states by Gaussian measurements~\cite{WisemanPRL2007}. The criterion~\eqref{eq:Reid} is also able to detect the steering of some non-Gaussian states~\cite{TehPRA2016,ReidJPA2017}, but its ability to capture complex distributions is ultimately limited by only considering first and second moments. The metrological approach thus provides particular advantages for the highly challenging problem of steering detection in non-Gaussian quantum states. This is in analogy to the metrological detection of entanglement that is known to be significantly more efficient in terms of the QFI instead of Gaussian quantifiers such as spin squeezing coefficients~\cite{PezzePRL2009,StrobelSCIENCE2014,PezzeRMP2018,GessnerPRL2019}.\\

\textbf{Bounds for specific measurements.}
Experimental tests of the condition~(\ref{eq:main}) are possible even without knowledge of the measurement settings that achieve the optimisations in Eqs.~(\ref{eq:assisted_qfi}) and~(\ref{eq:optcondvar}). Any fixed choice of local measurement settings $X$ and $X'$ for Alice and Bob, respectively, provides a joint sensitivity quantified by the (classical) Fisher information $\cfi^{\mathrm{AB}}[\assem_{\theta},X,X']$, and we obtain the hierarchy of inequalities (see Methods for a proof)
\begin{align}\label{eq:hierarchy}
\frac{1}{n\Vs{\theta_{\mathrm{est}}}}\leq \cfi^{\mathrm{AB}}[\assem_{\theta},X,X']\leq \qfi^{\mathrm{A},\mathrm{B}}[\assem_{\theta}]\leq \cqfi[\assem,H],
\end{align}
where $\qfi^{\mathrm{A},\mathrm{B}}[\assem_{\theta}]=\max_{X,X'}\cfi^{\mathrm{AB}}[\assem_{\theta},X,X']$ is the joint Fisher information, maximised over local measurement settings. Similarly, any fixed choice of $X$ yields an upper bound on~(\ref{eq:optcondvar}) and the inequalities 
\begin{align}\label{eq:Varhierarchy}
\cvar[\assem,H]\leq \sum_{a} p(a|X) \var[\rho^{\mathrm{B}}_{a|X},H] \leq \Vs{H_\mathrm{est}}
\end{align}
are saturated by an optimal measurement~(\ref{eq:optcondvar}) and estimator, respectively~\cite{ReidRMP2009}. These hierarchies reveal that any choice of local measurement settings leads to experimentally observable bounds for both sides of the inequality~(\ref{eq:main}). They further show how the simpler condition~(\ref{eq:steeringcondphasegen}) can be derived from~(\ref{eq:main}). Note that a different choice of setting $X$ must be used for estimating $\theta$ or $H$ in order to observe any effect from steering correlations. Both parties generally need to know which of the two settings is being used.\\

\textbf{Bounds on $\cqfi$ and $\cvar$.}
It is interesting to note that both sides of the inequality~(\ref{eq:main}) respect the same upper and lower bounds
\begin{align}\label{eq:upperlower}
\qfi[\rho^{\mathrm{B}},H]& \leq \cqfi[\assem,H] \stackrel{(*)}{\leq} 4 \var[\rho^{\mathrm{B}},H],\notag\\
\qfi[\rho^{\mathrm{B}},H] &\stackrel{(*)}{\leq}  4 \cvar[\assem,H] \leq 4 \var[\rho^{\mathrm{B}},H].
\end{align}
These inequalities hold for arbitrary assemblages $\assem$. 

When we can assume Alice's system to be quantum, we obtain the assemblage $\assem$ from the bipartite quantum state $\rho^{\mathrm{AB}}$ as $\assem(a,X) = \tr_{\mathrm{A}}[E^{\mathrm{A}}_{a|X} \rho^{\mathrm{AB}}]$, where the $E^{\mathrm{A}}_{a|X} \geq 0$ form a positive operator-valued measure (POVM) for the measurement setting $X$, normalised by $\sum_a E^{\mathrm{A}}_{a|X} = \id^{\mathrm{A}}$. The inequalities in~(\ref{eq:upperlower}) marked by $(*)$ are saturated when $\rho^{\mathrm{AB}}$ is a pure state, assuming Alice is able to perform any quantum measurement (see Methods). This result is a consequence of the remarkable facts that the QFI is the convex roof of the variance~\cite{Yu2013} while the variance is its own concave roof~\cite{TothPRA2013}, in addition to Alice being able to steer Bob's system into any pure-state ensemble for the local state $\rho^{\mathrm{B}}$.~\cite{HughstonPLA1993}

We construct explicit measurement bases for Alice to achieve steering in the optimal ensembles that saturate the above inequalities (Supplementary Note 5). We further observe that the inequality~\eqref{eq:main}, even with a fixed generator $H$, is capable of witnessing steering correlations for almost any pure state $\psi^{\mathrm{AB}}$. More precisely, \eqref{eq:main} is violated for any entangled $\psi^{\mathrm{AB}}$ whenever $H$ is not constant on the support of the local state $\rho^{\mathrm{B}}$.\\

\textbf{Steering of GHZ states.} Let us illustrate our criterion with a simple but relevant example. Consider a system composed of $N+1$ qubits, partitioned into a single control qubit (Alice) and the remaining $N$ qubits on Bob's side, that are prepared in a Greenberger-Horne-Zeilinger (GHZ) state of the form
\begin{equation}
	\ket{\mathrm{GHZ}_\phi^{N+1}} = \frac{1}{\sqrt{2}} \left( \ket{0}\ox {\ket{0}}^{\ox N} + e^{i\phi} \ket{1} \ox {\ket{1}}^{\ox N} \right),
\end{equation}
where $\ket{0},\ket{1}$ are eigenstates of the Pauli matrix $\sigma_z$. We take the local Hamiltonian $J^{\mathrm{B}}_z = \frac{1}{2}\sum_{i \in B} \sigma_z^{(i)}$, where the sum extends over the particles on Bob's side. When Alice measures her qubit in the $\sigma_z$ basis, Bob attains the quantum conditional variance $\cvar[\ket{\mathrm{GHZ}_\phi^{N+1}},J^{\mathrm{B}}_z] = 0$. GHZ states have the property~\cite{GessnerEPJQT2019} $\ket{\mathrm{GHZ}_\phi^{N+1}} = \frac{1}{\sqrt{2}} \left( \ket{+} \ox \ket{\mathrm{GHZ}_\phi^N} + \ket{-} \ox \ket{\mathrm{GHZ}_{\phi+\pi}^N} \right)$, where $\ket{+},\ket{-}$ are eigenstates of $\sigma_x$. This allows Alice to steer Bob's system into GHZ states by measuring in the $\sigma_x$ basis, and we obtain
\begin{align}
	&\quad\cqfi[\ket{\mathrm{GHZ}_\phi^{N+1}},J_z] \\& = \frac{1}{2} \left( \qfi[\ket{\mathrm{GHZ}_\phi^N},J^{\mathrm{B}}_z] + \qfi[\ket{\mathrm{GHZ}_{\phi+\pi}^N},J^{\mathrm{B}}_z] \right)  = N^2.\notag
\end{align}
This measurement is optimal and achieves the maximum in~(\ref{eq:assisted_qfi}) since $F_{\mathrm{Q}}[\rho,J^{\mathrm{B}}_z]\leq N^2$ holds for arbitrary quantum states~\cite{TothJPA2014,PezzeRMP2018}. Steering is detected by the clear violation of the condition~(\ref{eq:main}) for LHS models. So far the only known criteria able to detect steering in multipartite GHZ states are based on nonlocal observables that require individual addressing of the particles (see e.g. Refs.~\cite{CavalcantiPRA2011,ReidFRONTPHYS2012}), while our criterion is accessible by collective measurements. The criterion is moreover robust to white noise: For a mixture $\rho = p\ket{\mathrm{GHZ}_\phi^{N+1}}\bra{\mathrm{GHZ}_\phi^{N+1}} + (1-p)\id/2^{N+1}$, using the same measurements we obtain $\cqfi[\rho,J_z] \geq p^2 N^2/[p+2(1-p)/2^N],\, 4\cvar[\rho,J_z] \leq (1-p)N + p(1-p)N^2$. For large $N$, whenever $p \gtrapprox 1/\sqrt{N} $, the criterion witnesses steering. See Supplementary Note 2 for details and Supplementary Note 3 for an additional example involving a Schr\"odinger cat state.
\\

\textbf{Steering of atomic split twin Fock states}.
As an example of immediate practical relevance for state-of-the-art ultracold-atom experiments, consider $N/2$ spin excitations symmetrically distributed over $N$ particles, \ie a twin Fock state. Separating the particles into two addressable modes $\mathrm{A}$ and $\mathrm{B}$ with a $50:50$ beam splitter results in a split twin Fock state $|\mathrm{STF}_N\rangle$, which has been generated experimentally~\cite{LangeSCIENCE2018}. Similar experiments based on squeezed states were able to use Reid's criterion to verify steering~\cite{FadelSCIENCE2018,KunkelSCIENCE2018}, but the vanishing polarisation $\langle J^{\mathrm{B}}_x\rangle_{\rho^{\mathrm{B}}}=\langle J^{\mathrm{B}}_y\rangle_{\rho^{\mathrm{B}}}=0$ makes this challenging for split twin Fock states and so far only the entanglement between $A$ and $B$ could be detected~\cite{LangeSCIENCE2018}. We show that the criterion~(\ref{eq:main}) successfully reveals the EPR steering of split twin Fock states when Alice measures local spin observables $J_x^{\mathrm{A}}$, $J_z^{\mathrm{A}}$ and a phase shift $\theta$ is generated by $J_z^{\mathrm{B}}$. We obtain $\cvar[\ket{\mathrm{STF}_N},J_z^{\mathrm{B}}]=0$ and $\cqfi[\ket{\mathrm{STF}_N},J_z^{\mathrm{B}}]=N/4$, leading to a violation of~(\ref{eq:main}) that scales linearly with $N$. This value is limited by the partition noise that is introduced by the beam splitter which generates binomial fluctuations of the particle number in each mode. 

To overcome this limit, we propose the following alternative preparation of split Dicke states. Consider two addressable groups of $N/2$ atoms each. A collective measurement of the total number $k$ of spin excitations projects the system into a split Dicke state $|\mathrm{SD}_{N,k}\rangle$ without partition noise. This can be realised, e.g., with arrays of cold atoms in a cavity~\cite{HaasSCIENCE2014}. Using the same settings for Alice and Bob as before, these states still yield $\cvar[\ket{\mathrm{SD}_{N,k}},J_z^{\mathrm{B}}]=0$ while leading to significantly larger values of the quantum conditional Fisher information, and for the twin Fock case, $k=N/2$, we obtain the quadratic scaling $\cqfi[\ket{\mathrm{SD}_{N,N/2}},J_z^{\mathrm{B}}]=N(N+4)/12$; see Fig.~\ref{fig:2}. For details on arbitrary split Dicke states with and without partition noise, see Supplementary Note 4.\\

\begin{figure}
\centering
\includegraphics[width=.45\textwidth]{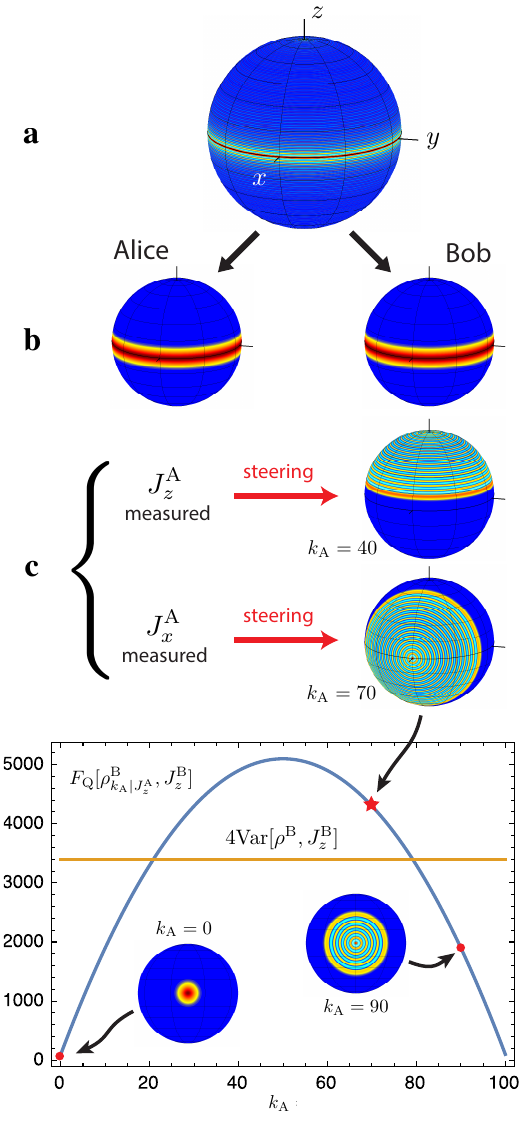}
\caption{\textbf{EPR-assisted metrology with twin Fock states.} \textbf{a)} We consider a twin Fock state with $N=200$ particles, that is split into two parts with $N_{\mathrm{A}}=N_{\mathrm{B}}=N/2$, here represented by the Wigner function on the Bloch sphere. \textbf{b)} The reduced state on either side is a mixture of Dicke states, resulting from tracing out the other half of the system. \textbf{c)} The two subsystems show perfect correlations for both measurement settings $J_x$ and $J_z$: When Alice measures $J_z^{\mathrm{A}}$ ($J_x^{\mathrm{A}}$) and obtains the result $k_{\mathrm{A}}$, she steers Bob's system into an eigenstate of $J_z^{\mathrm{B}}$ ($J_x^{\mathrm{B}}$) with eigenvalue $N/2-k_{\mathrm{A}}$. This can be used for assisted quantum metrology, and to reveal an EPR paradox. In the plot we show Bob's  sensitivity $\qfi[\rho^{\mathrm{B}}_{k_{\mathrm{A}}|J_x^{\mathrm{A}}},J_z^{\mathrm{B}}]$ when Alice obtains the result $k_{\mathrm{A}}$ from measuring $J_x^{\mathrm{A}}$ (blue line). Alice's results are all equally probable with $p(k_{\mathrm{A}}|J_x^{\mathrm{A}})=2/(N+2)$. Bob's average sensitivity $\cqfi[\ket{\mathrm{SD}_{N,N/2}},J_z^{\mathrm{B}}]$ coincides with the variance for the reduced state $4\var[\rho^{\mathrm{B}},J_z^{\mathrm{B}}]$ (yellow line), indicating that the measurement is optimal (Supplementary Note 4).}
\label{fig:2}
\end{figure}

\textbf{Phase estimation with multiple generators.}
Our result reveals the role of steering for generating probe states that are highly sensitive to the evolutions generated by a family of non-commuting generators, $\mathbf{H}=(H_1,\dots,H_m)$. We focus on a sequential scenario, where in each experimental trial a single parameter is generated by one of the elements of $\mathbf{H}$. Bob's estimation of the phase is assisted by steering from Alice who picks different measurement settings $X_i$ as a function of the acting Hamiltonian $H_i$ and includes details of $H_i$ in her communication to Bob. Achieving high sensitivity for multiple generators is relevant for multiparameter quantum metrology~\cite{HumphreysPRL2013,GessnerPRL2018,ProctorPRL2018,GePRL2018,GuoNatPhys2020,GessnerNatCommun2020}, but the identification of a single measurement observable that is suitable for all parameters~\cite{MatsumotoJPA2002,PezzePRL2017b} provides an additional complication that is not considered in our scenario.

A suitable figure of merit for Bob's average sensitivity is
\begin{align} \label{eq:multiparam}
\overline{\cqfi[\assem,\mathbf{H}]} = \sum_{i=1}^m\cqfi[\assem,H_i].
\end{align}
Using the same techniques as for the main inequality \eqref{eq:main}, we find that any assemblage admitting a LHS model satisfies (see Methods)
\begin{equation}
	\overline{\cqfi[\assem,\mathbf{H}]} \leq \max_{{\ket{\phi}}^{\mathrm{B}}} \sum_{i=1}^m 4\var[ {\proj{\phi}}^{\mathrm{B}}, H_i].
\end{equation}
An advantage over \eqref{eq:main} is that the right-hand side is state-independent. For a system $B$ of dimension $d$, we can take the $H_i$ to be a set of $d^2-1$ Hilbert-Schmidt orthonormal generators of $\mathrm{SU}(d)$,\cite{} and this bound simplifies to
\begin{equation} \label{eq:multiparam_finite}
	\overline{\cqfi[\assem,\mathbf{H}]} \leq 4(d-1).
\end{equation}
As a simple example, when Bob has a qubit ($d=2$), we can take the Pauli matrices as generators, $H_i = \sigma_i/\sqrt{2}, \, i=x,y,z$. Then \eqref{eq:multiparam_finite} becomes $\overline{\cqfi[\assem,\mathbf{H}]} \leq 4$. For a shared maximally entangled state, this inequality is violated since $\overline{\cqfi[\assem,\mathbf{H}]} = 6$. To interpret these numbers, note that any pure qubit state on Bob's side is optimal for sensing rotations about two orthogonal axes (each of which contributes a QFI of 2), but useless for the remaining axis. With a maximally entangled state, Alice can choose to steer Bob's system into a state that is optimal for whichever axis has been chosen, thus sensing about any given axis is optimal. \\

\textbf{Steering quantification. }
We may expect that the degree of violation of~\eqref{eq:main}, in a suitable sense, measures the amount of steering correlations. The proposed resource theory of steering~\cite{AolitaPRX2015} gives a set of criteria to be satisfied by a valid measure of steering. A general \emph{steering monotone} $\mc{S}$ assigns a non-negative real number to each assemblage. Firstly, we require (i) $\mc{S}(\assem) = 0$ for every assemblage $\assem$ with a LHV model. Next, (ii) $\mc{S}$ must be non-increasing (on average) when $\assem$ is operated upon by local operations and one-way classical communication from Bob to Alice (1W-LOCC). Finally, we may optionally require (iii) convexity: $\mc{S}(p \assem_1 + [1-p] \assem_2) \leq p \mc{S}(\assem_1)+ (1-p)\mc{S}(\assem_2)$ for any pair of assemblages classically mixed with probability $p$. Here, we propose two potential quantifiers and address whether they satisfy these criteria.

One quantity is the maximum possible violation of~\eqref{eq:main}, given the ability to vary the generator $H$. Since a rescaling of $H \to rH$ scales the QFI and the variance by the same factor $r^2$, we fix the norm of $H$ -- a convenient choice is to take $\tr[H^2] = 1$. Then the maximum violation of \eqref{eq:main} is
\begin{equation}
	\steerMax(\assem) := \max_{H,\, \tr[H^2]=1} \pos{ \frac{1}{4}\cqfi[\assem,H] - \cvar[\assem,H] },
\end{equation}
where $\pos{x} = \max\{0,x\}$. For a bipartite pure quantum state $\psi^{\mathrm{AB}}$, we have the easily computable formula (Supplementary Note 6) $\steerMax(\psi^{\mathrm{AB}}) = \lambda_\mathrm{max} [ \mathrm{diag}(\mathbf{p}) - \mathbf{p}\mathbf{p}^T]$,
where $\mathbf{p}$ is the vector of eigenvalues of $\rho^{\mathrm{B}}$ (equivalently, the Schmidt coefficients of $\psi^{\mathrm{AB}}$) and $\lambda_\mathrm{max}$ denotes the largest eigenvalue.

Alternatively, we can average over all $H$ with $\tr[H^2]=1$. Formally, this (rescaled) average is defined by
\begin{equation}
    \steerAvg(\assem) = (d^2-1) \pos{ \int \mu(\dd \mathbf{n}) \, \frac{1}{4} \cqfi[\assem,\mathbf{n}\cdot\mathbf{H}] - \cvar[\assem,\mathbf{n}\cdot\mathbf{H}]},
\end{equation}
where $H_i$ is any basis of orthonormal SU$(d)$ generators, and $\mu$ is the uniform measure over the sphere of unit vectors $\abs{\mathbf{n}}=1$. For pure states, we have $\steerAvg(\psi^{\mathrm{AB}}) =  \sum_{i\neq j} p_i p_j \left( 1 + \frac{2}{p_i+p_j} \right)$.

It follows immediately from~\eqref{eq:main} that both $\steerMax$ and $\steerAvg$ satisfy criterion (i). Moreover, we find that both are faithful indicators for pure states, meaning that they each vanish if and only if $\psi^{\mathrm{AB}}$ is separable. Convexity is also straightforward to prove (see Supplementary Note 6 for all details). Criterion (ii) can be ruled out for $\steerMax$ by again considering pure states: in this case, steering correlations (as with all correlations) are equivalent to entanglement~\cite{WisemanPRL2007,GisinPLA1991}. $\steerMax$ is found not to be an entanglement monotone; nevertheless, it remains an important quantity to consider if one is interested in observing the maximum possible violation. On the other hand, we prove that $\steerAvg$ is a pure state entanglement monotone. Thus it remains an open question whether $\steerAvg$ is in general a steering monotone.

\section*{Discussion}
We formulated the EPR paradox in the framework of quantum metrology, showing that it can be interpreted as an apparent violation of the complementarity relation between a local phase shift and its generator. This idea allowed us to derive a criterion to detect EPR correlations which is based on the quantum Fisher information, and thus stronger than known criteria based on the Heisenberg uncertainty relation. We illustrated this with concrete examples of non-Gaussian states in optical and atomic systems that are of immediate interest for experimental studies. By expressing the EPR paradox as a metrological task, our results demonstrate that such correlations can be useful for quantum-enhanced measurement protocols, thus having the potential to enable new sensing applications in quantum technologies.

\section*{Methods}
\textbf{Fisher information}. 
For a probability distribution $p(x|\theta)$ parameterised by $\theta \in \mathbb{R}$, the classical Fisher information is $\fisher[p(x|\theta)] := \int \dd x \, p(x|\theta) \left[ \partial_\theta \ln p(x|\theta) \right]^2$.
The quantum version $\qfi[\rho_\theta]$ for a parameter-dependent state $\rho_\theta$ may be defined as the maximum classical Fisher information associated with statistics obtained from any possible POVM $\{E_x\}$ via $p(x|\theta) = \tr[\rho_\theta E_x]$~\cite{BraunsteinPRL1994}. In the case of unitary parameter encoding $\rho_\theta = e^{-i\theta H} \rho e^{i\theta H}$ with a fixed generator $H$, the QFI is independent of $\theta$, so we denote it by $\qfi[\rho,H]$. This can be computed from the eigenvectors $\ket{\psi_i}$ and eigenvalues $\lambda_i$ of $\rho$:
\begin{equation}
    \qfi[\rho,H] = 2 \sum_{i,j \colon \lambda_i+\lambda_j\neq 0} \frac{(\lambda_i-\lambda_j)^2}{\lambda_i+\lambda_j} \abs{\braXket{\psi_i}{H}{\psi_j}}^2.
\end{equation}

\textbf{Proof of the main result}. Suppose $\assem$ is described by a LHS model, then
\begin{align}
\cqfi[\assem,H] & = \max_X \, \sum_a p(a|X) \qfi\left[ \sum_{\lambda} \frac{p(a|X,\lambda)p(\lambda)}{p(a|X)} \sigma^{\mathrm{B}}_\lambda, H \right] \nonumber\\
& \leq \max_X \, \sum_{a}\sum_\lambda p(a|X,\lambda)p(\lambda) \qfi[\sigma^{\mathrm{B}}_\lambda, H]\notag\\
& = \sum_{\lambda}p(\lambda)\qfi[\sigma^{\mathrm{B}}_\lambda, H],
\end{align}
where we used the convexity of the QFI~\cite{} and $\sum_ap(a|X,\lambda)=1$, since $\lambda$ and $X$ are independent. Making use of the upper bound~\cite{BraunsteinPRL1994,PezzeRMP2018} $\qfi[\rho, H]\leq 4 \var[\rho, H]$ that holds for arbitrary states $\rho$, we obtain
\begin{align}\label{eq:qfiLHS}
\cqfi[\assem,H] & \leq 4\sum_{\lambda}p(\lambda)\var[\sigma^{\mathrm{B}}_\lambda, H].
\end{align}
Moreover, following analogous steps, we obtain from the concavity of the variance~\cite{ReidRMP2009}
\begin{align}\label{eq:varLHS}
\cvar[\assem,H] \geq  \sum_{\lambda}p(\lambda)\var[\sigma^{\mathrm{B}}_\lambda, H].
\end{align}
Inserting~(\ref{eq:varLHS}) into~(\ref{eq:qfiLHS}) proves the result~(\ref{eq:main}).\\

\textbf{Recovering Reid's criterion}. The QFI describes the sensitivity for a parameter $\theta$ generated by $H$ that is achievable with an optimal measurement and estimation strategy. By using a specific estimator, constructed from the expectation value of some observable $M$, one obtains the lower bound \cite{PezzePRL2009,PezzeRMP2018}
\begin{align} \label{eq:qfi_lwr_var}
\qfi[\rho,H] \geq \frac{\abs{ \expect{[H,M]}_\rho }^2}{\var[\rho,M]}.
\end{align}
Together with the Cauchy-Schwarz inequality, we obtain for all $\assem$
\begin{align}\label{eq:lwrbndmethods}
\cqfi[\assem,H]
& \geq \max_X \, \sum_a p(a|X) \frac{\abs{\expect{[H,M]}_{\rho^{\mathrm{B}}_{a|X}}}^2}{\var[\rho^{\mathrm{B}}_{a|X},M]} \nonumber \\
& \geq \max_X \, \frac{\abs{ \sum_a p(a|X) \expect{[H,M]}_{\rho^{\mathrm{B}}_{a|X}} }^2}{ \sum_a p(a|X) \var[\rho^{\mathrm{B}}_{a|X},M]} \nonumber \\
& = \max_X \frac{ \abs{\expect{[H,M]}_{\rho^{\mathrm{B}}}}^2 }{\sum_a p(a|X) \var[\rho^{\mathrm{B}}_{a|X},M]} \nonumber \\
& = \frac{ \abs{\expect{[H,M]}_{\rho^{\mathrm{B}}}}^2 }{\cvar[\assem,M]}.
\end{align}
Inserting~(\ref{eq:lwrbndmethods}) into~(\ref{eq:main}) yields Reid's criterion. The formulation~(\ref{eq:Reid}) follows by using that $\Vs{M_{\mathrm{est}}}\geq \cvar[\assem,M]$ for all $M$.

In the case of a Gaussian quantum bipartite state with Gaussian measurements by Alice, Eq.~\eqref{eq:lwrbndmethods} can be saturated. Firstly, note that the quadrature variances (in fact, the whole covariance matrix) are identical for each conditional state $\rho_{a|X}^\mathrm{B}$\cite{FiurasekPRL2002}. A suitable pair of conjugate quadratures $H,M$ can be chosen such that Eq.~\eqref{eq:qfi_lwr_var} is saturated\cite{GessnerPRL2019}, thus the first inequality in~\eqref{eq:lwrbndmethods} is saturated. For the second inequality, note that all variances in the denominator are identical.

We can also directly recover Reid's criterion from the weaker condition~(\ref{eq:steeringcondphasegen}) by constructing an specific estimator from the measurement data $b,m$ of Alice and Bob, respectively. We assume that the dependence of the average value $\langle M_{\mathrm{est}}-M\rangle_{\theta}=\sum_b\sum_mp(b,m|Y,M,\theta)(m_{\mathrm{est}}(b)-m)$ on $\theta$ is known from calibration, where $p(b,m|Y,M,\theta)=p(b|Y)\langle m|\rho^{\mathrm{B}}_{b|Y,\theta}|m\rangle$. Given a sample of $n$ measurement results, the value of $\theta$ can now be estimated as the one that yields $\langle M_{\mathrm{est}}-M\rangle_{\theta}=\frac{1}{n}\sum_{i=1}^n(m_{\mathrm{est}}(b_i)-m_i)$. Without loss of generality we calibrate the estimator around the fixed value $\theta=0$, such that the estimator for $m$ is unbiased, \ie $\langle M_{\mathrm{est}}\rangle=\langle M\rangle_{\theta=0}$ (any biased estimator would lead to a larger error). The sample average evaluated at $\theta=0$ has a variance of $\frac{1}{n}\var[M_{\mathrm{est}}]$. Note that only the distribution of Bob's results $m_i$ depends on $\theta$, and therefore $|\frac{\partial}{\partial \theta}\langle M_{\mathrm{est}}-M\rangle_{\theta}|=|\frac{\partial}{\partial \theta}\langle M\rangle_{\theta}|$. In the central limit ($n\gg 1$), this strategy therefore yields a sensitivity of
\begin{align}
    \var[\theta_{\mathrm{est}}] = \frac{\var[M_{\mathrm{est}}]}{n\left|\frac{\partial \langle M\rangle_{\theta}}{\partial \theta}\right|^2} \;,
\end{align}
which can be shown from a maximal likelihood analysis of the sample average distribution or from Gaussian error propagation~\cite{Varenna}. We obtain the result Eq.~(\ref{eq:variancebound}). \\

\textbf{Sensitivity for fixed local measurements}. For fixed measurement settings $X$ and $X'$, respectively, the joint statistics of Alice and Bob are described by the probability distribution $p(a,b|X,X',\theta)=p(a|X)\mathrm{Tr}\{E_{b|X'}\rho^{\mathrm{B}}_{a|X,\theta}\}$ where $E_{b|X'}$ is a positive operator-valued measure (POVM) describing the measurement $X'$. The Cram\'er-Rao bound
\begin{align}
n\Vs{\theta_{\mathrm{est}}}\geq 1/\cfi^{\mathrm{AB}}[\assem_{\theta},X,X']
\end{align}
identifies the precision limit for any estimator that is constructed from the local measurement results $a$ and $b$ and for any choice of $X$ and $X'$ in terms of the Fisher information
\begin{align}
\cfi^{\mathrm{AB}}[\assem_{\theta},X,X'] = \sum_{a,b}p(a,b|X,X',\theta)\left(\frac{\partial}{\partial \theta}\log p(a,b|X,X',\theta)\right)^2.
\end{align}
A straightforward calculation reveals that
\begin{align}
\cfi^{\mathrm{AB}}[\assem_{\theta},X,X'] = \sum_a p(a|X) \cfi^{\mathrm{B}}[X'|\rho^{\mathrm{B}}_{a|X,\theta}],
\end{align}
\ie for fixed settings, the joint sensitivity coincides with Bob's average conditional sensitivity $\cfi^{\mathrm{B}}[X'|\rho^{\mathrm{B}}_{a|X,\theta}]=\sum_b\mathrm{Tr}\{E_{b|X'}\rho^{\mathrm{B}}_{a|X,\theta}\}\left(\frac{\partial}{\partial \theta}\log \mathrm{Tr}\{E_{b|X'}\rho^{\mathrm{B}}_{a|X,\theta}\}\right)^2$ since Alice's data is independent of $\theta$. Maximising over the choice of measurement yields the hierarchy
\begin{align}
\cfi^{\mathrm{AB}}[\assem_{\theta},X,X']  &\leq \underbrace{\max_X\max_{X'}\sum_a p(a|X) \cfi^{\mathrm{B}}[X'|\rho^{\mathrm{B}}_{a|X,\theta}]}_{\qfi^{\mathrm{A},\mathrm{B}}[\assem_{\theta}]}\notag\\
&\leq \max_X\sum_a p(a|X)\underbrace{\max_{X'} \cfi^{\mathrm{B}}[X'|\rho^{\mathrm{B}}_{a|X,\theta}]}_{F_{\mathrm{Q}}[\rho^{\mathrm{B}}_{a|X},H]}\notag\\
&=\cqfi[\assem,H].
\end{align}
This completes the proof for the set of inequalities~(\ref{eq:hierarchy}).\\

\textbf{Metrological steering for bipartite quantum states}. Let us first note that if Alice's system is quantum, the optimal measurements in~(\ref{eq:optcondvar}) and~(\ref{eq:assisted_qfi}) can always be implemented by rank-$1$ POVMs. This follows from the convexity of the QFI and the concavity of the variance (Supplementary Note 1).

Now suppose that $\rho^{\mathrm{AB}}$ is pure. Since the optimal POVM for $\cqfi$ is rank-1, the corresponding conditional states $\rho^{\mathrm{B}}_{a|X}$ are pure. An important fact about bipartite pure states is that any pure-state ensemble on Bob's side (consistent with the average state $\rho^{\mathrm{B}}$) may be realised by an appropriate rank-1 POVM on Alice's side~\cite{HughstonPLA1993}. Thus the optimisation can be reduced to
\begin{align} \label{eqn:cqfi_concave_roof}
\cqfi[\rho^{\mathrm{AB}},H] & = \max_{\substack{ \{p(a),\, \ket{\phi_a} \}_a \colon \\ \sum_a p(a) \proj{\phi_a} = \rho^{\mathrm{B}}}}  \, \sum_a p(a) \qfi[\proj{\phi_a}, H] \nonumber \\
& = \max_{\substack{ \{p(a),\, \ket{\phi_a} \}_a \colon \\ \sum_a p(a) \proj{\phi_a} = \rho^{\mathrm{B}}}} \, 4\sum_a p(a) \var[\proj{\phi_a}, H]\notag\\
&=4\var[\rho^{\mathrm{B}},H].
\end{align}
In the last line we used that the variance is its own concave roof~\cite{TothPRA2013}. For $\cvar$ the minimisation is the same as taking the convex roof, resulting in~\cite{Yu2013} $\qfi[\rho^{\mathrm{B}},H]$. Hence, for a pure state $\rho^{\mathrm{AB}}$, we obtain the equalities $\cqfi[\rho^{\mathrm{AB}},H]=4\var[\rho^{\mathrm{B}},H]$ and $4\cvar[\rho^{\mathrm{AB}},H]=\qfi[\rho^{\mathrm{B}},H]$. For arbitrary assemblages, we obtain the upper bounds $\cqfi[\assem,H]\leq4\var[\rho^{\mathrm{B}},H]$ and $4\cvar[\assem,H]\geq\qfi[\rho^{\mathrm{B}},H]$ as a consequence of convexity of the QFI, concavity of the variance, and $\qfi[\rho,H] \leq 4\var[\rho,H]$. For the same reason, we obtain that $\cqfi[\assem,H]\geq \qfi[\rho^{\mathrm{B}},H]$ and $\cvar[\assem,H]\leq \var[\rho^{\mathrm{B}},H]$ for arbitrary assemblages $\assem$, including those obtained from $\rho^{\mathrm{AB}}$. This concludes the proof of~(\ref{eq:upperlower}).\\

\textbf{Multiple generators}.
One can ask whether there is a (potentially weaker) steering witness involving only the QFI. It is clear that the right-hand side of \eqref{eq:main} cannot be made state-independent: the best one can do is to replace $\cvar[\assem,H]$ by $\max_{\sigma} \var[\sigma,H]$, leading to an inequality that holds for all cases, even non-steerable.

Instead, we turn to the quantity \eqref{eq:multiparam}. Without any assistance from Alice, the best achievable precision would be
\begin{equation}
	\overline{\qfi[\rho^{\mathrm{B}}, \mathbf{H}]} := \sum_{i=1}^m \qfi[\rho^{\mathrm{B}},H_i].
\end{equation}
Following the same technique as for a single parameter, any LHS model satisfies
\begin{align}
	\overline{\cqfi[\assem,\mathbf{H}]} & \leq \overline{\qfi^*[\mathbf{H}]} \nonumber \\
		& := \max_{\sigma^{\mathrm{B}}} \, \overline{\qfi[\sigma^{\mathrm{B}},\mathbf{H}]} \nonumber \\
		& = \max_{{\ket{\phi}}^{\mathrm{B}}} \, \overline{\qfi[{\proj{\phi}}^{\mathrm{B}}, \mathbf{H}]} \nonumber \\
		& = \max_{{\ket{\phi}}^{\mathrm{B}}} \, \sum_i 4 \var[{\proj{\phi}}^{\mathrm{B}}, H_i].
\end{align}
The fact that pure states achieve the maximum on the right-hand side follows from convexity of the QFI. This bound is of course only possible when the $H_i$ are bounded.

Using the same techniques as for $\steerAvg$ (Supplementary Note 6), we can take $H_i$ to be a set of $d^2-1$ traceless generators of $\mathrm{SU}(d)$ satisfying $\tr[H_i H_j]=\delta_{i,j}$, and compute $\overline{\qfi^*[\mathbf{H}]} = \overline{\qfi[{\proj{\phi}}^{\mathrm{B}}, \mathbf{H}]} = 4(d-1)$ (which actually holds for any $\ket{\phi}$). Thus, for this set of $\mathbf{H}$ in $d$ dimensions, the LHS bound is
\begin{equation} \label{eqn:discrete_simple_bound}
	\overline{\qfi[\assem,\mathbf{H}]} \leq 4(d-1).
\end{equation}
For a pure state $\psi^{\mathrm{AB}}$, 
\begin{align}
	\overline{\cqfi[\psi^{\mathrm{AB}},\mathbf{H}]} & = \sum_i 4\var[\rho^{\mathrm{B}}, H_i] \nonumber \\
		& = 4(d-1) + 4 \sum_{i\neq j} p_i p_j,
\end{align}
so that \eqref{eqn:discrete_simple_bound} is violated if and only if $\psi^{\mathrm{AB}}$ is entangled.

\textbf{Acknowledgments}\\
We thank G. Adesso, L. Pezz\`e, A. Smerzi, and P. Treutlein for discussions. 
BY acknowledges financial support from the European Research Council (ERC) under the Starting Grant GQCOP (Grant No.~637352) and grant number (FQXi FFF Grant number FQXi-RFP-1812) from the Foundational Questions Institute and Fetzer Franklin Fund, a donor advised fund of Silicon Valley Community Foundation.
MF was partially supported by the Swiss National Science Foundation, and by the Research Fund of the University of Basel for Excellent Junior Researchers.
MG acknowledges funding by the LabEx ENS-ICFP: ANR-10-LABX-0010/ANR-10-IDEX-0001-02 PSL*.\\

\textbf{Author contributions}\\
BY, MF and MG contributed equally to this work.

\textbf{Data availability}\\
All relevant data are available from the authors.

\textbf{Code availability}\\
Source codes of the plots are available from the authors upon request.

\textbf{Competing interests}\\
The authors declare no competing interests.

\clearpage
\newpage
\onecolumngrid

\section*{Supplementary Information}

\section*{Supplementary Note 1 - Optimal POVMs for assisted metrology}\label{sec:optPOVM}
Here, we argue that the optimal measurements performed by Alice can always be taken as rank-1 POVMs, for any assemblage defined by a global quantum state. Suppose a rank-$r$ POVM $E^{\mathrm{A}}_{a|X}$ is optimal for the conditional QFI -- that is,
\begin{equation}
    \cqfi[\rho^{\mathrm{AB}},H] = \sum_a p(a|X) \qfi[\rho^{\mathrm{B}}_{a|X},H].
\end{equation}
Then we can decompose (for instance, using the spectral decomposition) $E^{\mathrm{A}}_{a|X} = \sum_{i=1}^r E^{\mathrm{A}}_{a,i|X}$, where each $E^{\mathrm{A}}_{a,i|X} \geq 0$ is at most rank-1, This defines a new, fine-grained POVM with conditional states $p(a,i|X) = \tr_{\mathrm{A}}[ E^{\mathrm{A}}_{a,i|X} \rho^{\mathrm{AB}}]$. The original conditional states are obtained by averaging over $i$: $p(a|X) \rho^{\mathrm{B}}_{a|X} = \sum_i \tr_{\mathrm{A}}[E^{\mathrm{A}}_{a,i|X} \rho^{\mathrm{AB}}] = \sum_i p(a,i|X) \rho^{\mathrm{B}}_{a,i|X}$. Due to convexity of the QFI,
\begin{align}
    \sum_{a,i} p(a,i|X) \qfi[\rho^{\mathrm{B}}_{a,i|X},H] & = \sum_a p(a|X) \sum_i \frac{p(a,i|X)}{p(a|X)} \qfi[ \rho^{\mathrm{B}}_{a_i|X},H] \nonumber \\
        & \geq \sum_a p(a|X) \qfi \left[ \sum_i \frac{p(a,i|X)}{p(a|X)} \rho^{\mathrm{B}}_{a,i|X}, H \right] \nonumber \\
        & = \sum_a p(a|X) \qfi[ \rho^{\mathrm{B}}_{a|X}, H] \nonumber \\
        & = \cqfi[\rho^{\mathrm{AB}}, H]. 
\end{align}
Thus the fine-grained POVM is also optimal. The same conclusion holds for the quantum conditional variance, instead using concavity of the variance and the fact that the optimal POVM must minimise the average variance.

\section*{Supplementary Note 2 - GHZ states with white noise}
We first observe that, for any pure state $\psi$ mixed with white noise in $d$ dimensions~\cite{TothJPA2014},
\begin{equation} \label{eq:qfi_white_noise}
    \qfi\left[ p\psi + \frac{(1-p)}{d} \id, H\right] = \frac{4p^2}{p+2(1-p)/d} \var[\psi,H].
\end{equation}
This follows from choosing an eigenbasis $\ket{i},\, i=0,\dots,d-1$ for the mixed state with $\ket{0}=\ket{\psi}$ and expanding in terms of its eigenvalues $\lambda_i$:~\cite{Paris2009}
\begin{align}
    \qfi\left[ p\psi + \frac{(1-p)}{d} \id, H\right] & = 4 \sum_{i<j} \frac{(\lambda_i-\lambda_j)^2}{\lambda_i+\lambda_j} \abs{\braXket{i}{H}{j}}^2 \nonumber \\
        & = 4 \sum_{j>0} \frac{p^2}{p+2(1-p)/d} \bra{0}H \proj{j} H \ket{0} \nonumber \\
        & = \frac{4 p^2}{p+2(1-p)/d} \bra{\psi}H(I - \proj{\psi})H\ket{\psi} \nonumber \\
        & = \frac{4p^2}{p+2(1-p)/d} \left[ \braXket{\psi}{H^2}{\psi} - \braXket{\psi}{H}{\psi}^2 \right].
\end{align}
For the shared GHZ state $\ket{\mathrm{GHZ}_\phi^{N+1}} = \frac{1}{\sqrt{2}} \left( \ket{0}\ox {\ket{0}}^{\ox N} + e^{i\phi} \ket{1} \ox {\ket{1}}^{\ox N} \right)$ mixed with white noise, any projection onto a pure state on Alice's side results in the same conditional state as obtained for the pure case, up to a mixture with the identity on Bob's side. We keep the same measurement choices for any $p$, although they may not be optimal when $p<1$.

For a measurement of $\sigma_z$ by Alice, Bob's conditional states are easily found to give
\begin{equation}
    \cvar[\rho, J_z] \leq \frac{(1-p)N}{4} + \frac{p(1-p)N^2}{4}.
\end{equation}
With a $\sigma_x$ measurement, \eqref{eq:qfi_white_noise} results in
\begin{equation}
    \cqfi[\rho, J_z] \geq \frac{p^2N^2}{p + 2(1-p)/d},
\end{equation}
where $d=2^N$. When $p \gg 1/d = 2^{-N}$, we can neglect the term involving $d$. Then $\cqfi[\rho,J_z] \gtrapprox p N^2$, and the difference
\begin{equation}
    \cqfi[\rho, J_z] - 4\cvar[\rho, J_z] \gtrapprox p^2 N^2 - (1-p)N
\end{equation}
is positive as long as $N > (1-p)/p^2$. For large $N$, this condition approximates to $p \gtrapprox 1/\sqrt{N}$.

\section*{Supplementary Note 3 - Hybrid cat states}

\begin{figure}[b]
\centering
\includegraphics[width=.5\textwidth]{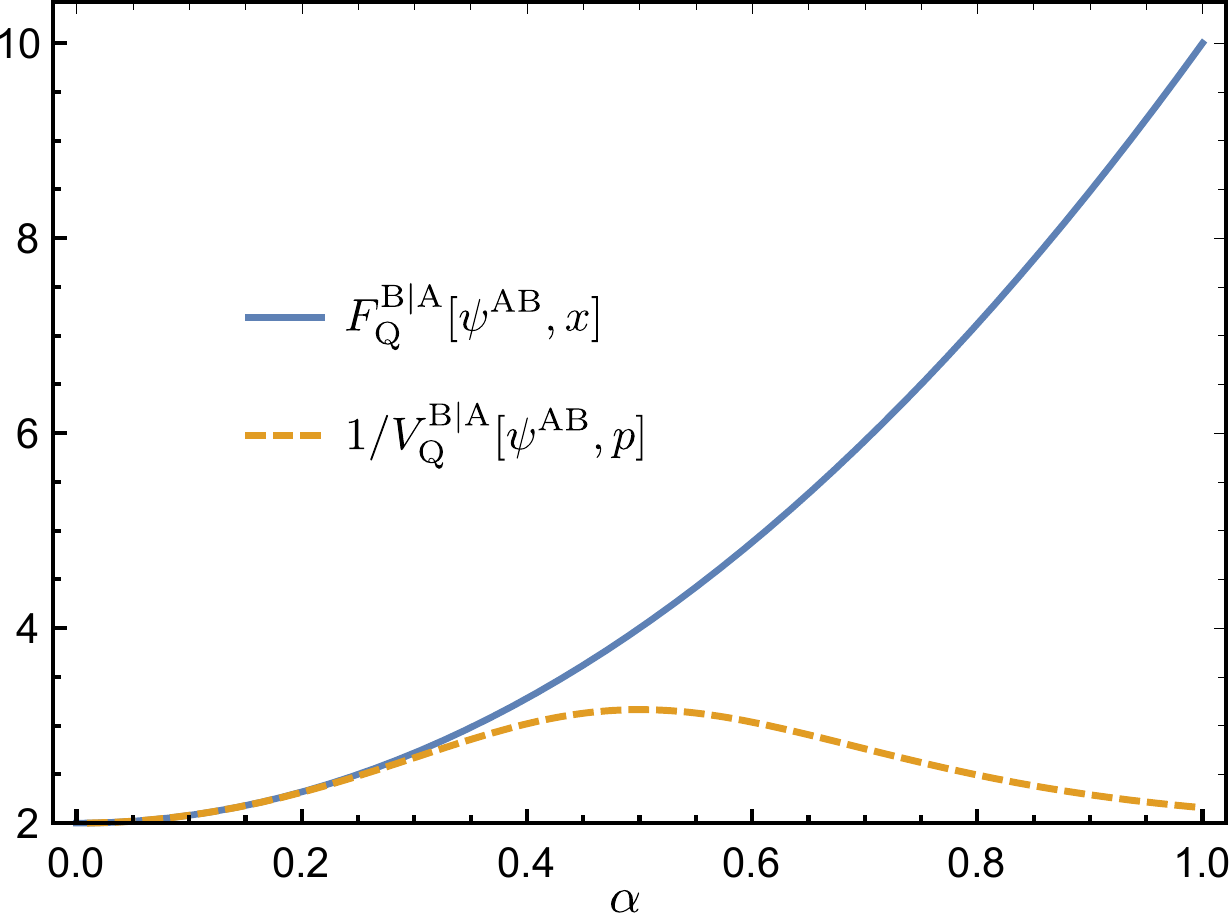}
\caption{\textbf{Hybrid cat state.} Alice has a qubit and Bob has a bosonic mode, entangled as in Supplementary Eq.~(\ref{eq:hybridcat}). Plot comparing the the conditional QFI $\cqfi[\psi^\mathrm{AB},x]$ (blue solid) against the lower bound $1/\cvar[\psi^\mathrm{AB},p]$ (orange dashed), as a function of the coherent state parameter $\alpha$.
}
\label{fig:cat}
\end{figure}

Here, we consider an example of a hybrid system where Alice has a qubit and Bob has a single bosonic mode. Consider the bipartite ``cat state"
\begin{equation}\label{eq:hybridcat}
    {\ket{\psi}}_{\mathrm{AB}} = \frac{1}{\sqrt{2}} \left( {\ket{0}}_{\mathrm{A}} {\ket{\alpha}}_{\mathrm{B}} + {\ket{1}}_{\mathrm{A}} {\ket{-\alpha}}_{\mathrm{B}} \right),
\end{equation}
where $\ket{\pm \alpha}$ are coherent states and we take $\alpha \geq 0$ and use a quadrature observable $H = x$.

First consider projection by Alice onto a pure state $\ket{\chi} = a\ket{0} + b\ket{1}$, resulting in Bob's conditional state $\ket{\phi} = (a^*\ket{\alpha} + b^*\ket{-\alpha})/\sqrt{2q}$ with probability $q = 1/2 + \Re[ab^*]e^{-2\alpha^2}$.
Representing $\ket{\chi}$ in terms of the unit Bloch vector $\mathbf{r} = (x,y,z)^T$, one finds $q = (1+x e^{-2\alpha^2})/2$ and the expectation values $q \braXket{\phi}{x}{\phi} = \alpha z/\sqrt{2}$, $q\braXket{\phi}{x^2}{\phi} = \alpha^2 + q/2$, so the variance is
\begin{align}
    q \var[\proj{\phi},x] & = \alpha^2 \left[1 - \frac{z^2}{2q}\right] + \frac{q}{2}.
\end{align}
Now let Alice use an arbitrary rank-1 POVM with elements $E_m = 2 k_m \proj{\chi_m}, m=0,1,\dots,M,\, k_m \geq 0$ (which can be assumed without loss of generality -- see Supplementary Section~\ref{sec:optPOVM}).
In terms of unit Bloch vectors $\mathbf{r}_m$, these can be expressed as
\begin{equation}
    E_m = k_m (\id + \mathbf{r}_m \cdot \boldsymbol{\sigma})
\end{equation}
The completeness condition $\sum_m E_k = \id$ is then equivalent to
\begin{equation}
    \sum_m k_m = 1, \quad \sum_m k_m \mathbf{r}_m = \mathbf{0}.
\end{equation}
The probability of outcome $m$ is $p_m = 2k_m q_m$, where $q_m = \| {\bra{\chi_m}}_\mathrm{A} {\ket{\psi}}_\mathrm{AB}\|^2$.
Hence the average variance conditional on this POVM is 
\begin{align}
    \sum_m p_m \var[\proj{\phi_m},x] & = \sum_m \left[ 2k_m \alpha^2\left( 1 - \frac{z_m^2}{2q_m} \right) + \frac{p_m}{2} \right]\\
        & = 2\alpha^2 \left[ 1 - \sum_m \frac{k_m z_m^2}{1+ x_m e^{-2\alpha^2}}\right] + \frac{1}{2}.
\end{align}
From the fact that $z_m^2/(1+x_m e^{-2\alpha^2}) \in [0,1]$, it follows that $\sum_m k_m z_m^2/(1+x_m e^{-2\alpha^2}) \in [0,1]$, and so
\begin{equation}
    \frac{1}{2} \leq \sum_m p_m \var[\proj{\phi_m}, x] \leq 2\alpha^2 + \frac{1}{2}.
\end{equation}
The lower bound is saturated by using a measurement in the $\{\ket{0},\ket{1}\}$ basis, giving
\begin{equation}
    \cvar[\psi^{\mathrm{AB}}, x] = \frac{1}{2}.
\end{equation}
On the other hand, the upper bound is saturated with any POVM whose coordinates $z_m$ all vanish -- i.e., all the Bloch vectors lie on the equator -- for example the $\{\ket{+},\ket{-}\}$ basis, giving
\begin{equation}
    \frac{1}{4} \cqfi[\psi^{\mathrm{AB}}, x] = 2\alpha^2 + \frac{1}{2}.
\end{equation}
Hence we see that steering is witnessed via this strategy for any nonzero $\alpha$, and the violation increases with $\alpha$.

This may be compared with the Reid criterion using $x$ and $p$ quadrature variances~\cite{Reid2008}. From Eq.~(25) in the main text, a lower bound on the conditional QFI is $\cqfi[\psi^{\mathrm{AB}},x] \geq 1/\cvar[\psi^{\mathrm{AB}},p]$. Using the same techniques as above, we find $\cvar[\psi^{\mathrm{AB}},p] = 1/2 - 2\alpha^2 e^{-4\alpha^2}$, obtained using a measurement in the basis $(\ket{0}\pm i \ket{1})/\sqrt{2}$. While steering is witnessed for all nonzero $\alpha$, the violation is greatest at $\alpha=1/2$ and the criterion becomes less effective as $\alpha$ increases -- see Supplementary Fig.~\ref{fig:cat}.

\section*{Supplementary Note 4 - Atomic split Dicke states}
In this Section, we apply our criterion to detect steering between two addressable atomic ensembles with a fixed number of total excitations. We first consider in~\ref{sec:SDnopartnoise} the deterministic distribution of $N$ atoms in two modes with a fixed number of excitations, as proposed in the main text of our manuscript. Then, in~\ref{sec:SDpartnoise} we analyse a Dicke state that is sent onto a spatial beam splitter to separate each of its mode in two, as was done experimentally with an ensemble of $N=5\,000$ atoms in Ref.~\cite{LangeSCIENCE2018}.

\subsection{Dicke states with fixed splitting $N_{\mathrm{A}}:N_{\mathrm{B}}$}\label{sec:SDnopartnoise}
Consider $N$ atoms split into two addressable modes A and B, with respectively $N_{\mathrm{A}}$ and $N_{\mathrm{B}}=N-N_{\mathrm{A}}$ particles. Assume that we know that the internal spin degree of freedom of a total number of $0\leq k \leq N$ atoms is excited (e.g., from a collective measurement), but we do not know the distribution of the excited atoms into the two modes. The system is described by the split Dicke state
\begin{align}\label{stex1}
|\mathrm{SD}_{k,N_{\mathrm{A}}:N_{\mathrm{B}}}\rangle=\mathcal{N}\sum_{\substack{k_{\mathrm{A}},k_{\mathrm{B}}\\k_{\mathrm{A}}+k_{\mathrm{B}}=k}}|k_{\mathrm{A}}\rangle\otimes|k_{\mathrm{B}}\rangle \;,
\end{align}
where we introduced the eigenstates with $k_X$ excitations of the $N_X$-particle spin observable $J_z^{X}$, for $X=\mathrm{A},\mathrm{B}$,
\begin{align}
J_z^X|k_X\rangle=(k_X-N_X/2)|k_X\rangle \;.
\end{align}
The range of $k_{\mathrm{A}}$ and $k_{\mathrm{B}}$ in the sum depends on the values of $k$, $N_{\mathrm{A}}$ and $N_{\mathrm{B}}$. We can formulate constaints, e.g., in terms of $k_{\mathrm{A}}$: Since (i) if there are $k>N_{\mathrm{B}}$ excitations in total, the number of excitations in A must be at least $k_{\mathrm{A}}=k-N_{\mathrm{B}}$ and (ii) $N_{\mathrm{A}}$ atoms can show at most $N_{\mathrm{A}}$ excitations. These constraints can be taken into account explicitly as
\begin{align}\label{eq:splitDickenopnoise}
|\mathrm{SD}_{k,N_{\mathrm{A}}:N_{\mathrm{B}}}\rangle=\frac{1}{\sqrt{k_{\max}-k_{\min}+1}}\sum_{k_{\mathrm{A}}=k_{\min}}^{k_{\max}}|k_{\mathrm{A}}\rangle\otimes|k-k_{\mathrm{A}}\rangle \;,
\end{align}
where
\begin{align}\label{eq:kminmax}
k_{\min}&=\max\{0,k-N_{\mathrm{B}}\}\notag\\
k_{\max}&=\min\{k,N_{\mathrm{A}}\} \;.
\end{align}

\subsubsection{Alice measures $J_z^{\mathrm{A}}$, Bob measures $J_z^{\mathrm{A}}$}
To determine the conditional variance, we consider the projection of Alice's system (A) onto eigenstates $|k_{\mathrm{A}}\rangle$ of $J_z^{\mathrm{A}}$. Alice obtains any of the results $k_{\mathrm{A}}=k_{\min},\dots,k_{\max}$ with probability $p(k_{\mathrm{A}}|J_z^{\mathrm{A}})=1/(k_{\max}-k_{\min}+1)$, while other results have probability zero. Bob's conditional state $|\Psi_{k_{\mathrm{A}}|J_z^{\mathrm{A}}}\rangle=|k-k_{\mathrm{A}}\rangle$ has zero variance for $J_z^{\mathrm{B}}$, and we obtain
\begin{align}\label{eqFix:cqv}
\cvar[|\mathrm{SD}_{k,N_{\mathrm{A}}:N_{\mathrm{B}}}\rangle,J_z^{\mathrm{B}}]=\sum_{k_{\mathrm{A}}}p(k_{\mathrm{A}}|J_z^{\mathrm{A}})\var[|k-k_{\mathrm{A}}\rangle,J_z^{\mathrm{B}}]=0 \;.
\end{align}
This measurement is therefore optimal in the sense that it achieves the minimum in the definition of $\cvar[|\mathrm{SD}_{k,N_{\mathrm{A}}:N_{\mathrm{B}}}\rangle,J_z^{\mathrm{B}}]$ [see Eq.~(2) in the main text]. This result can be understood intuitively: knowing the total number $k$ of excitations along with the fact that $k_{\mathrm{A}}$ of them are found in Alice's subsystem, allows us to predict with certainty that Bob will measure $k_{\mathrm{B}}=k-k_{\mathrm{A}}$ excitations.

\subsubsection{Alice measures $J_x^{\mathrm{A}}$, Bob estimates $\theta$}
For the estimation of a phase shift $\theta$ generated by $J_z^{\mathrm{B}}$ on Bob's subsystem, let us now consider the measurement of $J_x^{\mathrm{A}}$ by Alice, described by projection onto the eigenstates $|k_{\mathrm{A}}\rangle_x=e^{-i\frac{\pi}{2}J_y^{\mathrm{A}}}|k_{\mathrm{A}}\rangle$. The assemblage is given by
\begin{align}
\assem(k_{\mathrm{A}},J_x^{\mathrm{A}})&=\mathrm{Tr}_{\mathrm{A}}\{(|k_{\mathrm{A}}\rangle_x\langle k_{\mathrm{A}}|_x\otimes\id)|\mathrm{SD}_{k,N_{\mathrm{A}}:N_{\mathrm{B}}}\rangle\langle\mathrm{SD}_{k,N_{\mathrm{A}}:N_{\mathrm{B}}}|\}\notag\\
&=p(k_{\mathrm{A}}|J_x^{\mathrm{A}})|\Psi_{k_{\mathrm{A}}|J_x^{\mathrm{A}}}\rangle\langle\Psi_{k_{\mathrm{A}}|J_x^{\mathrm{A}}}|,
\end{align}
with conditional states
\begin{align}\label{ragu}
|\Psi_{k_{\mathrm{A}}|J_x^{\mathrm{A}}}\rangle&=\frac{1}{\sqrt{\sum_{k'_{\mathrm{A}}=k_{\min}}^{k_{\max}}|\langle k_{\mathrm{A}}|e^{i\frac{\pi}{2}J_y^{\mathrm{A}}}|k'_{\mathrm{A}}\rangle|^2}}\sum_{k'_{\mathrm{A}}=k_{\min}}^{k_{\max}}\langle k_{\mathrm{A}}|e^{i\frac{\pi}{2}J_y^{\mathrm{A}}}|k'_{\mathrm{A}}\rangle|k-k'_{\mathrm{A}}\rangle,
\end{align}
and probabilities
\begin{align}
p(k_{\mathrm{A}}|J_x^{\mathrm{A}})=\frac{1}{k_{\max}-k_{\min}+1}\sum_{k'_{\mathrm{A}}=k_{\min}}^{k_{\max}}|\langle k_{\mathrm{A}}|e^{i\frac{\pi}{2}J_y^{\mathrm{A}}}|k'_{\mathrm{A}}\rangle|^2.
\end{align}

The overlap between eigenstates of $J_z^{\mathrm{A}}$ and $J_x^{\mathrm{A}}$ can be computed using the expression
\begin{align}\label{eq:zxoverlap}
\langle k_{\mathrm{A}} \vert e^{-i \phi J_y^{\mathrm{A}}} \vert k_{\mathrm{A}}^\prime \rangle = \sqrt{k_{\mathrm{A}}^\prime!(N_{\mathrm{A}}-k_{\mathrm{A}}^\prime)!k_{\mathrm{A}}!(N_{\mathrm{A}}-k_{\mathrm{A}})!} \sum_{n=\max[k_{\mathrm{A}}^\prime-k_{\mathrm{A}},0]}^{\min[k_{\mathrm{A}}^\prime,N_{\mathrm{A}}-k_{\mathrm{A}}]} \dfrac{(-1)^n\cos(\phi/2)^{k_{\mathrm{A}}-k_{\mathrm{A}}^\prime+N_{\mathrm{A}}-2n}\sin(\phi/2)^{2n+k_{\mathrm{A}}^\prime-k_{\mathrm{A}}}}{(k_{\mathrm{A}}-n)!(N_{\mathrm{A}}-k_{\mathrm{A}}^\prime-n)!n!(k_{\mathrm{A}}^\prime-k_{\mathrm{A}}+n)!} \;.
\end{align}

We obtain the first and second moments of the conditional states
\begin{align}
\langle J_z^{\mathrm{B}}\rangle_{k_{\mathrm{A}}|J_x^{\mathrm{A}}}&=\frac{1}{\sum_{k'_{\mathrm{A}}=k_{\min}}^{k_{\max}}|\langle k_{\mathrm{A}}|e^{i\frac{\pi}{2}J_y^{\mathrm{A}}}|k'_{\mathrm{A}}\rangle|^2}\sum_{k'_{\mathrm{A}}=k_{\min}}^{k_{\max}}\vert \bra{k_{\mathrm{A}}} e^{i\frac{\pi}{2}J_y^{\mathrm{A}}} \ket{k_{\mathrm{A}}^\prime} \vert^2  \bra{k-k_{\mathrm{A}}^\prime} J_z^{\mathrm{B}} \ket{k-k_{\mathrm{A}}^\prime} \notag\\
&=\frac{1}{\sum_{k'_{\mathrm{A}}=k_{\min}}^{k_{\max}}|\langle k_{\mathrm{A}}|e^{i\frac{\pi}{2}J_y^{\mathrm{A}}}|k'_{\mathrm{A}}\rangle|^2}\sum_{k'_{\mathrm{A}}=k_{\min}}^{k_{\max}}\vert \bra{k_{\mathrm{A}}} e^{i\frac{\pi}{2}J_y^{\mathrm{A}}} \ket{k_{\mathrm{A}}^\prime} \vert^2  \left(k - k_{\mathrm{A}}^\prime-\dfrac{N_{\mathrm{B}}}{2} \right) \label{amatriciana}
\end{align}
and
\begin{align}
\langle (J_z^{\mathrm{B}})^2\rangle_{k_{\mathrm{A}}|J_x^{\mathrm{A}}}
&=\frac{1}{\sum_{k'_{\mathrm{A}}=k_{\min}}^{k_{\max}}|\langle k_{\mathrm{A}}|e^{i\frac{\pi}{2}J_y^{\mathrm{A}}}|k'_{\mathrm{A}}\rangle|^2}\sum_{k'_{\mathrm{A}}=k_{\min}}^{k_{\max}}\vert \bra{k_{\mathrm{A}}} e^{i\frac{\pi}{2}J_y^{\mathrm{A}}} \ket{k_{\mathrm{A}}^\prime} \vert^2  \left(k - k_{\mathrm{A}}^\prime-\dfrac{N_{\mathrm{B}}}{2} \right)^2.
\label{pesto}
\end{align}
Since the conditional states are pure, this yields a quantum Fisher information of
\begin{align}
\qfi[|\Psi_{k_{\mathrm{A}}|J_x^{\mathrm{A}}}\rangle,J_z^{\mathrm{B}}]=4\var[|\Psi_{k_{\mathrm{A}}|J_x^{\mathrm{A}}}\rangle,J_z^{\mathrm{B}}]&=4(\langle (J_z^{\mathrm{B}})^2\rangle_{k_{\mathrm{A}}|J_x^{\mathrm{A}}}-\langle J_z^{\mathrm{B}}\rangle_{k_{\mathrm{A}}|J_x^{\mathrm{A}}}^2).
\end{align}
Generally, any choice of Alice's measurement setting $X$ yields a lower bound for the quantum conditional Fisher information:
\begin{align}\label{eq:specificconditionalFisherinfo}
    \cfi^{\mathrm{B}|\mathrm{A}}[\assem,X,H]:=\sum_ap(a|X)\qfi[\rho^{\mathrm{B}}_{a|X},H]\leq \max_X\cfi^{\mathrm{B}|\mathrm{A}}[\assem,X,H]=\cqfi[\assem,H].
\end{align}
We obtain the conditional Fisher information
\begin{align}\label{eqFix:cqfi}
\cfi^{\mathrm{B}|\mathrm{A}}[|\mathrm{SD}_{k,N_{\mathrm{A}}:N_{\mathrm{B}}}\rangle,J_x^{\mathrm{A}},J_z^{\mathrm{B}}]= 4\sum_{k_{\mathrm{A}}=0}^{N_{\mathrm{A}}}p(k_{\mathrm{A}}|J_x^{\mathrm{A}})(\langle (J_z^{\mathrm{B}})^2\rangle_{k_{\mathrm{A}}|J_x^{\mathrm{A}}}-\langle J_z^{\mathrm{B}}\rangle_{k_{\mathrm{A}}|J_x^{\mathrm{A}}}^2).
\end{align}

\subsubsection{Reduced quantum Fisher information and variance}
The properties of Bob's reduced state provide upper and lower limits on the quantum conditional variance and quantum conditional Fisher information, respectively; cf. Eq.~(11) in the main text. Bob's reduced density matrix is given as
\begin{align}
\rho^{\mathrm{B}}=\frac{1}{k_{\max}-k_{\min}+1}\sum_{k_{\mathrm{A}}=k_{\min}}^{k_{\max}} |k-k_{\mathrm{A}}\rangle\langle k-k_{\mathrm{A}}|.
\end{align}
Let us first calculate the variance of $J_z^{\mathrm{B}}$. The first moment reads
\begin{align}
\langle J_z^{\mathrm{B}}\rangle_{\rho^{\mathrm{B}}}&=\frac{1}{k_{\max}-k_{\min}+1}\sum_{k_{\mathrm{A}}=k_{\min}}^{k_{\max}}\bra{k-k_{\mathrm{A}}} J_z^{\mathrm{B}} \ket{k-k_{\mathrm{A}}}\notag\\
&=\frac{1}{k_{\max}-k_{\min}+1}\sum_{k_{\mathrm{A}}=k_{\min}}^{k_{\max}} \left(k - k_{\mathrm{A}}-\dfrac{N_{\mathrm{B}}}{2} \right)\notag\\
&=\frac{1}{2} \left(2 k - N_{\mathrm{B}} - k_{\min} - k_{\max}\right)
\end{align}
and for the second moments, we obtain
\begin{align}
\langle (J_z^{\mathrm{B}})^2\rangle_{\rho^{\mathrm{B}}}&=\frac{1}{k_{\max}-k_{\min}+1}\sum_{k_{\mathrm{A}}=k_{\min}}^{k_{\max}} \left(k - k_{\mathrm{A}}-\dfrac{N_{\mathrm{B}}}{2} \right)^2\notag\\
&=\frac{1}{12} \left(3 (N_{\mathrm{B}}-2 k)^2 + 4 k_{\min}^2 + 
   2 k_{\max} (3 N_{\mathrm{B}} - 6 k + 2 k_{\max}+1) + 
   2 k_{\min}(3 N_{\mathrm{B}} - 6k+ 2 k_{\max}-1)\right)
\end{align}
We obtain
\begin{align}\label{eqFix:rqv}
\var[\rho^{\mathrm{B}},J_z^{\mathrm{B}}]=\frac{1}{12} (k_{\max}-k_{\min} +2) (k_{\max}-k_{\min}).
\end{align}

Since the state $\rho^{\mathrm{B}}$ is invariant under transformations generated by $J_z^{\mathrm{B}}$, \ie $[\rho^{\mathrm{B}},J_z^{\mathrm{B}}]=0$, we obtain that the quantum Fisher information of Bob's reduced state vanishes, $\qfi[\rho^{\mathrm{B}},J_z^{\mathrm{B}}]=0$. This can be confirmed explicitly using the expression
\begin{align}\label{eq:mixedQFI}
\qfi[\rho^{\mathrm{B}},J_z^{\mathrm{B}}]=2\sum_{i,j}\frac{(p_i-p_{j})^2}{p_i+p_{j}}|\langle\psi_i|J_z^{\mathrm{B}}|\psi_j\rangle|^2,
\end{align}
where $\rho^{\mathrm{B}}=\sum_{i}p_i|\psi_i\rangle\langle\psi_i|$ is the spectral decomposition of $\rho^{\mathrm{B}}$ with eigenvalues $p_i=1/(k_{\max}-k_{\min}+1)$ and eigenvectors $|\psi_i\rangle =|k-i\rangle$.

\begin{figure}[tb]
\centering
\includegraphics[width=.8\textwidth]{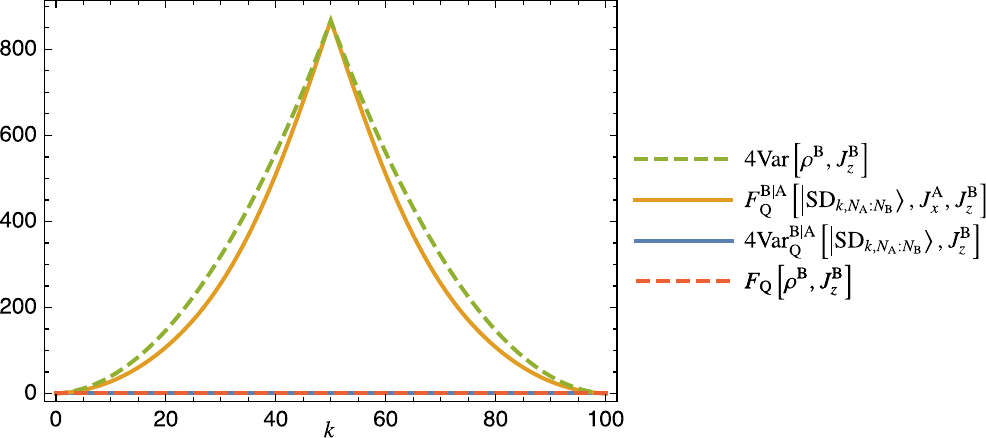}
\caption{\textbf{Split Dicke state without partition noise.} Dicke state with $N=100$ particles and $k$ excitations, deterministically split into $N_{\mathrm{A}}=N_{\mathrm{B}}=N/2$. Plot of Supplementary Eq.~\eqref{eqFix:cqv} (blue), Eq.~\eqref{eqFix:cqfi} (yellow), Eq.~\eqref{eqFix:rqv} (green dashed) and Eq.~\eqref{eq:mixedQFI} (red dashed).
}
\label{fig:SDkfix}
\end{figure}

\subsubsection{Results for a twin Fock state divided into $N_{\mathrm{A}}=N_{\mathrm{B}}=N/2$}
When the initial state is a twin Fock state, \ie $k=N/2$, that is split in two equal parts with $N_{\mathrm{A}}=N_{\mathrm{B}}=N/2$, the above expressions simplifies further. First of all, we obtain $k_{\min}=0$, $k_{\max}=N/2$, and $\sum_{k'_{\mathrm{A}}=k_{\min}}^{k_{\max}}|\langle k_{\mathrm{A}}|e^{i\frac{\pi}{2}J_y^{\mathrm{A}}}|k'_{\mathrm{A}}\rangle|^2 = 1$ (as the sum runs over the full basis of the $N_{\mathrm{A}}=N/2$ particle state), giving for the probabilities $p(k_{\mathrm{A}}|J_x^{\mathrm{A}})=2/(N+2)$. To simplify the conditional states~(\ref{ragu}), note from~(\ref{eq:zxoverlap}) that $\langle k_{\mathrm{A}}|e^{-iJ^{\mathrm{A}}_y\phi}|k_{\mathrm{A}}'\rangle=\langle N_{\mathrm{A}}-k_{\mathrm{A}}|e^{-iJ^{\mathrm{A}}_y\phi}|N_{\mathrm{A}}-k_{\mathrm{A}}'\rangle$ and that $\langle k_{\mathrm{A}}|e^{-iJ^{\mathrm{A}}_y\phi}|k_{\mathrm{A}}'\rangle=\langle k_{\mathrm{A}}'|e^{iJ^{\mathrm{A}}_y\phi}|k_{\mathrm{A}}\rangle$. Moreover, the matrix elements of $J^{\mathrm{A}}_y$ and $J^{\mathrm{B}}_y$ coincide since both operators are of the same length. We use this in Supplementary Eq.~(\ref{ragu}) to write
\begin{align}\label{eq:condX}
    |\Psi_{k_{\mathrm{A}}|J_x^{\mathrm{A}}}\rangle&=\sum_{k'_{\mathrm{A}}=0}^{N/2}\langle k_{\mathrm{A}}|e^{i\frac{\pi}{2}J_y^{\mathrm{A}}}|k'_{\mathrm{A}}\rangle|N/2-k'_{\mathrm{A}}\rangle\notag\\
    &=\sum_{k'_{\mathrm{A}}=0}^{N/2}\langle k'_{\mathrm{A}}|e^{-i\frac{\pi}{2}J_y^{\mathrm{A}}}|k_{\mathrm{A}}\rangle|N/2-k'_{\mathrm{A}}\rangle\notag\\
    &=\sum_{k'_{\mathrm{A}}=0}^{N/2}\langle N/2- k'_{\mathrm{A}}|e^{-i\frac{\pi}{2}J_y^{\mathrm{A}}}|N/2-k_{\mathrm{A}}\rangle|N/2-k'_{\mathrm{A}}\rangle\notag\\
    &=\sum_{k'_{\mathrm{A}}=0}^{N/2}\langle N/2- k'_{\mathrm{A}}|e^{-i\frac{\pi}{2}J_y^{\mathrm{B}}}|N/2-k_{\mathrm{A}}\rangle|N/2-k'_{\mathrm{A}}\rangle\notag\\
    &=\underbrace{\left(\sum_{k'_{\mathrm{A}}=0}^{N/2}|N/2-k'_{\mathrm{A}}\rangle\langle N/2- k'_{\mathrm{A}}|\right)}_{\id_{\mathrm{B}}}e^{-i\frac{\pi}{2}J_y^{\mathrm{B}}}|N/2-k_{\mathrm{A}}\rangle\notag\\
    &=|N/2-k_{\mathrm{A}}\rangle_x.
\end{align}
Hence, just like in the early examples by EPR and Bohm\cite{EPR1935,ReidRMP2009}, the state shows perfect correlations in two non-commuting measurement bases, and we may express~(\ref{eq:splitDickenopnoise}) as
\begin{align}
    |\mathrm{SD}_{\frac{N}{2},\frac{N}{2}:\frac{N}{2}}\rangle&=\sqrt{\frac{2}{N+2}}\sum_{k_{\mathrm{A}}=0}^{N/2}|k_{\mathrm{A}}\rangle\otimes|N/2-k_{\mathrm{A}}\rangle\notag\\
    &=\sqrt{\frac{2}{N+2}}\sum_{k_{\mathrm{A}}=0}^{N/2}|k_{\mathrm{A}}\rangle_x\otimes|N/2-k_{\mathrm{A}}\rangle_x.
\end{align}
Using~(\ref{eq:condX}), we determine the first and second moments of $J_z^{\mathrm{B}}$ for the conditional states to be
\begin{align}
\langle J_z^{\mathrm{B}}\rangle_{k_{\mathrm{A}}|J_x^{\mathrm{A}}}&=0,
\end{align}
and
\begin{align}
\langle (J_z^{\mathrm{B}})^2\rangle_{k_{\mathrm{A}}|J_x^{\mathrm{A}}}=\frac{1}{8} (2 k_{\mathrm{A}} (N-2 k_{\mathrm{A}}) + N).
\end{align}
This leads to a quantum conditional Fisher information of 
\begin{align}
\cqfi[|\mathrm{SD}_{\frac{N}{2},\frac{N}{2}:\frac{N}{2}}\rangle,J_z^{\mathrm{B}}]&=4\sum_{k_{\mathrm{A}}=0}^{N/2}p(k_{\mathrm{A}}|J_x^{\mathrm{A}})(\langle (J_z^{\mathrm{B}})^2\rangle_{k_{\mathrm{A}}|J_x^{\mathrm{A}}}=\sum_{k_{\mathrm{A}}=0}^{N/2}\frac{1}{N+2}(2 k_{\mathrm{A}} (N-2 k_{\mathrm{A}}) + N)=\frac{1}{12} N(4 + N).
\end{align}
By comparison with Supplementary Eq.~(\ref{eqFix:rqv}), which yields $\var[\rho^{\mathrm{B}},J_z^{\mathrm{B}}] = \frac{1}{48} N(4 + N)$, we notice that the upper bound $\cqfi[|\mathrm{SD}_{k,N_{\mathrm{A}}:N_{\mathrm{B}}}\rangle,J_z^{\mathrm{B}}]= 4\var[\rho^{\mathrm{B}},J_z^{\mathrm{B}}]$ [see Eq.~(11) in the main text] is indeed saturated by this choice of measurement. This shows that no other measurement by Alice could yield a higher average sensitivity on Bob's side. The measurement of $J_x^{\mathrm{A}}$ is optimal for assisted metrology with split twin Fock states as it achieves the maximum in the definition of $\cqfi[|\mathrm{SD}_{\frac{N}{2},\frac{N}{2}:\frac{N}{2}}\rangle,J_z^{\mathrm{B}}]$ [see Eq.~(5) in the main text].

\subsection{Splitting a Dicke state into two modes}\label{sec:SDpartnoise}
We now focus on a preparation of split Dicke states by a beam splitter operation. Consider a Dicke state with $k$ excitations in the modes $a$ and $b$, described as
\begin{align}\label{eq:Dicke}
|\mathrm{D}_{k,N}\rangle=\frac{(a^{\dagger})^{k}(b^{\dagger})^{N-k}}{\sqrt{k!(N-k)!}}|0\rangle.
\end{align}
By sending this state onto a beam splitter with ratio $p:1-p$, both modes are split by into two modes as
\begin{align}\label{eq:splitter}
a^\dagger = \sqrt{p} a_{\mathrm{A}}^\dagger + \sqrt{1-p} a_{\mathrm{B}}^\dagger, \notag\\
b^\dagger = \sqrt{p} b_{\mathrm{A}}^\dagger + \sqrt{1-p} b_{\mathrm{B}}^\dagger.
\end{align}
As a consequence of the partition noise, the total number of particles in each mode fluctuates. By expanding the binomials that appear upon inserting~(\ref{eq:splitter}) into~(\ref{eq:Dicke}), this state can be written as
\begin{align}
|\mathrm{SD}_{k,N,p}\rangle&=\sum_{k_{\mathrm{A}}=0}^k\sum_{N_{\mathrm{A}}=k_{\mathrm{A}}}^{N-k+k_{\mathrm{A}}}\sqrt{\binom{k}{k_{\mathrm{A}}}\binom{N-k}{N_{\mathrm{A}}-k_{\mathrm{A}}}}\sqrt{p}^{N_{\mathrm{A}}}\sqrt{1-p}^{N-N_{\mathrm{A}}}|k_{\mathrm{A}}\rangle_{N_{\mathrm{A}}}\otimes|k-k_{\mathrm{A}}\rangle_{N-N_{\mathrm{A}}}\notag\\
&=\sum_{N_{\mathrm{A}}=0}^N\sum_{k_{\mathrm{A}}=k_{\min}}^{k_{\max}}\sqrt{\binom{k}{k_{\mathrm{A}}}\binom{N-k}{N_{\mathrm{A}}-k_{\mathrm{A}}}}\sqrt{p}^{N_{\mathrm{A}}}\sqrt{1-p}^{N-N_{\mathrm{A}}}|k_{\mathrm{A}}\rangle_{N_{\mathrm{A}}}\otimes|k-k_{\mathrm{A}}\rangle_{N-N_{\mathrm{A}}} \;,
\end{align}
where $|k_{\mathrm{A}}\rangle_{N_{\mathrm{A}}}$ is an eigenstate of the spin-$N_{\mathrm{A}}/2$ observable $J^{\mathrm{A}}_z$, and similarly for subsystem B.

Alice's measurements of $J_z^{\mathrm{A}}$ or $J_x^{\mathrm{A}}$, provide simultaneous information about the spin quantum number and the number of particles $N_{\mathrm{A}}$, whose observable commutes with all spin components. Typically, after a suitable rotation of the state, one measures how many spins point up/down, such that the information about the total number of particles is provided simultaneously. Alice could ignore the information provided by $N_{\mathrm{A}}$, but this coarse-graining would lead to sub-optimal results for the conditional variance and quantum Fisher information; see Supplementary Note~\ref{sec:optPOVM}. 

\subsubsection{Alice measures $J_z^{\mathrm{A}}$, Bob measures $J_z^{\mathrm{B}}$}
A measurement of $J_z^{\mathrm{A}}$ with the result $k_{\mathrm{A}}$ for the magnetic quantum number $k_{\mathrm{A}}$ and $N_{\mathrm{A}}$ for the number of particles occurs with probability
\begin{align}
p(k_{\mathrm{A}},N_{\mathrm{A}}|J_z^{\mathrm{A}})=\binom{k}{k_{\mathrm{A}}}\binom{N-k}{N_{\mathrm{A}}-k_{\mathrm{A}}}p^{N_{\mathrm{A}}}(1-p)^{N-N_{\mathrm{A}}}
\end{align}
for all $k_{\min}\leq k_{\mathrm{A}}\leq k_{\max}$ and with zero probability otherwise. This event produces the conditional state
\begin{align}
|\mathrm{SD}_{k,N,p}\rangle_{k_{\mathrm{A}},N_{\mathrm{A}}|J_z^{\mathrm{A}}}&=|k-k_{\mathrm{A}}\rangle_{N-N_{\mathrm{A}}}
\end{align}
on Bob's side. Since these are eigenstates of $J_z^{\mathrm{B}}$ we obtain that
\begin{align}\label{eqPN:cqv}
\cvar[|\mathrm{SD}_{k,N,p}\rangle,J_z^{\mathrm{B}}]=0 \;.
\end{align}
This reflects the fact that a measurement of $J_z^{\mathrm{A}}$ and $N_{\mathrm{A}}$ allows to predict with certainty the measurement results for $J_z^{\mathrm{B}}$ and $N_{\mathrm{B}}$.

\subsubsection{Alice measures $J_x^{\mathrm{A}}$, Bob estimates $\theta$}
For the estimation of a phase shift generated by $J_z^{\mathrm{A}}$, we consider measurements of $J_x^{\mathrm{A}}$, together with $N_{\mathrm{A}}$, with results $(k_{\mathrm{A}},N_{\mathrm{A}})$. A straightforward calculation shows that the event $(k_{\mathrm{A}},N_{\mathrm{A}})$ occurs with probability
\begin{align}
p(k_{\mathrm{A}},N_{\mathrm{A}}|J_x^{\mathrm{A}})=p^{N_{\mathrm{A}}}(1-p)^{N-N_{\mathrm{A}}}\sum_{k'_{\mathrm{A}}=k_{\min}}^{k_{\max}}\binom{k}{k'_{\mathrm{A}}}\binom{N-k}{N_{\mathrm{A}}-k'_{\mathrm{A}}}|\langle k_{\mathrm{A}}|e^{i\frac{\pi}{2}J_y^{\mathrm{A}}}|k'_{\mathrm{A}}\rangle|^2 \;,
\end{align}
for $0\leq N_{\mathrm{A}}\leq N$ and $0\leq k_{\mathrm{A}}\leq N_{\mathrm{A}}$. Bob's conditional state in this case reads
\begin{align}
|\mathrm{SD}_{k,N,p}\rangle_{k_{\mathrm{A}},N_{\mathrm{A}}|J_x^{\mathrm{A}}}&=\frac{1}{\sqrt{p(k_{\mathrm{A}},N_{\mathrm{A}}|J_x^{\mathrm{A}})}}\sum_{k'_{\mathrm{A}}=k_{\min}}^{k_{\max}}\sqrt{\binom{k}{k'_{\mathrm{A}}}\binom{N-k}{N_{\mathrm{A}}-k'_{\mathrm{A}}}}\sqrt{p}^{N_{\mathrm{A}}}\sqrt{1-p}^{N-N_{\mathrm{A}}}\langle k_{\mathrm{A}}|e^{i\frac{\pi}{2}J_y^{\mathrm{A}}}|k'_{\mathrm{A}}\rangle|k-k'_{\mathrm{A}}\rangle_{N-N_{\mathrm{A}}}\notag\\
&=\frac{1}{\sqrt{\sum_{k'_{\mathrm{A}}=k_{\min}}^{k_{\max}}\binom{k}{k'_{\mathrm{A}}}\binom{N-k}{N_{\mathrm{A}}-k'_{\mathrm{A}}}|\langle k_{\mathrm{A}}|e^{i\frac{\pi}{2}J_y^{\mathrm{A}}}|k'_{\mathrm{A}}\rangle|^2
}}\sum_{k'_{\mathrm{A}}=k_{\min}}^{k_{\max}}\sqrt{\binom{k}{k'_{\mathrm{A}}}\binom{N-k}{N_{\mathrm{A}}-k'_{\mathrm{A}}}}\langle k_{\mathrm{A}}|e^{i\frac{\pi}{2}J_y^{\mathrm{A}}}|k'_{\mathrm{A}}\rangle|k-k'_{\mathrm{A}}\rangle_{N-N_{\mathrm{A}}} \;.\label{eq:SkANASx}
\end{align}
These states have the expectation value
\begin{align}
\langle J_z^{\mathrm{B}}\rangle_{k_{\mathrm{A}},N_{\mathrm{A}}|J_x^{\mathrm{A}}}=\frac{1}{\sum_{k'_{\mathrm{A}}=k_{\min}}^{k_{\max}}\binom{k}{k'_{\mathrm{A}}}\binom{N-k}{N_{\mathrm{A}}-k'_{\mathrm{A}}}|\langle k_{\mathrm{A}}|e^{i\frac{\pi}{2}J_y^{\mathrm{A}}}|k'_{\mathrm{A}}\rangle|^2
}\sum_{k'_{\mathrm{A}}=k_{\min}}^{k_{\max}}\binom{k}{k'_{\mathrm{A}}}\binom{N-k}{N_{\mathrm{A}}-k'_{\mathrm{A}}}|\langle k_{\mathrm{A}}|e^{i\frac{\pi}{2}J_y^{\mathrm{A}}}|k'_{\mathrm{A}}\rangle|^2\left(k-k'_{\mathrm{A}}-\frac{N_{\mathrm{B}}}{2}\right) \;,
\end{align}
and second moment
\begin{align}
\langle (J_z^{\mathrm{B}})^2\rangle_{k_{\mathrm{A}},N_{\mathrm{A}}|J_x^{\mathrm{A}}}=\frac{1}{\sum_{k'_{\mathrm{A}}=k_{\min}}^{k_{\max}}\binom{k}{k'_{\mathrm{A}}}\binom{N-k}{N_{\mathrm{A}}-k'_{\mathrm{A}}}|\langle k_{\mathrm{A}}|e^{i\frac{\pi}{2}J_y^{\mathrm{A}}}|k'_{\mathrm{A}}\rangle|^2
}\sum_{k'_{\mathrm{A}}=k_{\min}}^{k_{\max}}\binom{k}{k'_{\mathrm{A}}}\binom{N-k}{N_{\mathrm{A}}-k'_{\mathrm{A}}}|\langle k_{\mathrm{A}}|e^{i\frac{\pi}{2}J_y^{\mathrm{A}}}|k'_{\mathrm{A}}\rangle|^2\left(k-k'_{\mathrm{A}}-\frac{N_{\mathrm{B}}}{2}\right)^2 \;,
\end{align}
yielding the quantum Fisher information
\begin{align}
\qfi[|\mathrm{SD}_{k,N,p}\rangle_{k_{\mathrm{A}},N_{\mathrm{A}}|J_x^{\mathrm{A}}},J_z^{\mathrm{B}}]=4\var[|\mathrm{SD}_{k,N,p}\rangle_{k_{\mathrm{A}},N_{\mathrm{A}}|J_x^{\mathrm{A}}},J_z^{\mathrm{B}}]=4(\langle (J_z^{\mathrm{B}})^2\rangle_{k_{\mathrm{A}},N_{\mathrm{A}}|J_x^{\mathrm{A}}}-\langle J_z^{\mathrm{B}}\rangle_{k_{\mathrm{A}},N_{\mathrm{A}}|J_x^{\mathrm{A}}}^2) \;.
\end{align}
This choice of measurements leads to the conditional Fisher information [see Supplementary Eq.~(\ref{eq:specificconditionalFisherinfo})]:
\begin{align}\label{eqPN:cqfi}
\cfi^{\mathrm{B}|\mathrm{A}}[|\mathrm{SD}_{k,N,p}\rangle,J_x^{\mathrm{A}},J_z^{\mathrm{B}}]= \sum_{k_{\mathrm{A}}=0}^k\sum_{N_{\mathrm{A}}=k_{\mathrm{A}}}^{N-k+k_{\mathrm{A}}}p(k_{\mathrm{A}},N_{\mathrm{A}}|J_x^{\mathrm{A}})\qfi[|\mathrm{SD}_{k,N,p}\rangle_{k_{\mathrm{A}},N_{\mathrm{A}}|J_x^{\mathrm{A}}},J_z^{\mathrm{B}}] \;.
\end{align}

\subsubsection{Reduced quantum Fisher information and variance}
Bob's reduced state is given by
\begin{align}
\rho^{\mathrm{B}}&=\mathrm{Tr}_{\mathrm{A}}\{|\mathrm{SD}_{k,N,p}\rangle\langle \mathrm{SD}_{k,N,p}|\}\\
&=\sum_{k_{\mathrm{A}}=0}^k\sum_{N_{\mathrm{A}}=k_{\mathrm{A}}}^{N-k+k_{\mathrm{A}}}\binom{k}{k_{\mathrm{A}}}\binom{N-k}{N_{\mathrm{A}}-k_{\mathrm{A}}}p^{N_{\mathrm{A}}}(1-p)^{N-N_{\mathrm{A}}}|k-k_{\mathrm{A}}\rangle_{N-N_{\mathrm{A}}}\langle k-k_{\mathrm{A}}|_{N-N_{\mathrm{A}}}\:.
\end{align}
It yields the average value
\begin{align}
\langle J_z^{\mathrm{B}}\rangle_{\rho^{\mathrm{B}}}&=\sum_{k_{\mathrm{A}}=0}^k\sum_{N_{\mathrm{A}}=k_{\mathrm{A}}}^{N-k+k_{\mathrm{A}}}\binom{k}{k_{\mathrm{A}}}\binom{N-k}{N_{\mathrm{A}}-k_{\mathrm{A}}}p^{N_{\mathrm{A}}}(1-p)^{N-N_{\mathrm{A}}}\langle k-k_{\mathrm{A}}|J_z^{\mathrm{B}}|k-k_{\mathrm{A}}\rangle_{N-N_{\mathrm{A}}}\notag\\
&=\sum_{k_{\mathrm{A}}=0}^k\sum_{N_{\mathrm{A}}=k_{\mathrm{A}}}^{N-k+k_{\mathrm{A}}}\binom{k}{k_{\mathrm{A}}}\binom{N-k}{N_{\mathrm{A}}-k_{\mathrm{A}}}p^{N_{\mathrm{A}}}(1-p)^{N-N_{\mathrm{A}}}\left(k-k_{\mathrm{A}}-\frac{N-N_{\mathrm{A}}}{2}\right)\notag\\
&=\left(\frac{N}{2}-k\right)(1-p).
\end{align}
For the second moment, we obtain
\begin{align}
\langle (J_z^{\mathrm{B}})^2\rangle_{\rho^{\mathrm{B}}}&=\sum_{k_{\mathrm{A}}=0}^k\sum_{N_{\mathrm{A}}=k_{\mathrm{A}}}^{N-k+k_{\mathrm{A}}}\binom{k}{k_{\mathrm{A}}}\binom{N-k}{N_{\mathrm{A}}-k_{\mathrm{A}}}p^{N_{\mathrm{A}}}(1-p)^{N-N_{\mathrm{A}}}\left(k-k_{\mathrm{A}}-\frac{N-N_{\mathrm{A}}}{2}\right)^2\notag\\
&=\left(\frac{N}{2}-k\right)^2(1-p)^2+\frac{N}{4}p(1-p),
\end{align}
and the variance reads
\begin{align}\label{eqPN:rqv}
\var[\rho^{\mathrm{B}},J_z^{\mathrm{B}}]=\frac{N}{4}p(1-p).
\end{align}

\begin{figure}[tb]
\centering
\includegraphics[width=.8\textwidth]{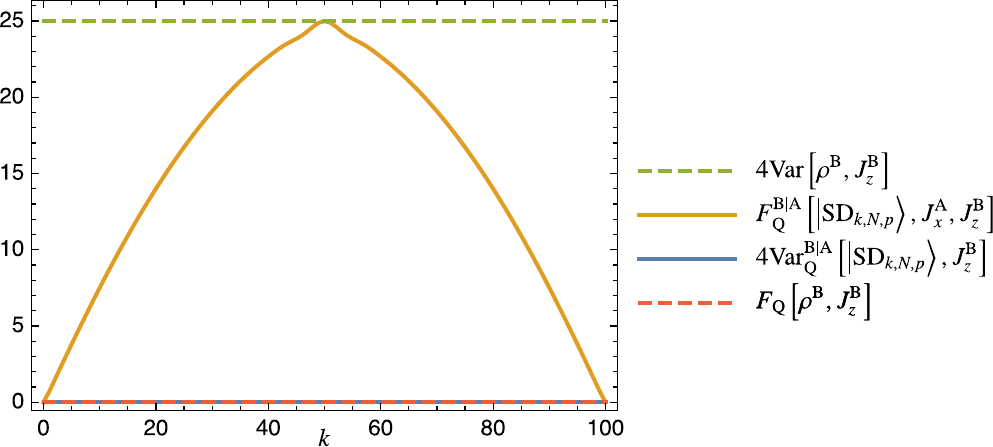}
\caption{\textbf{Dicke state split with partition noise.} Dicke state with $N=100$ particles and $k$ excitations, split into two modes with $50:50$ ratio. Plot of Supplementary Eq.~\eqref{eqPN:cqv} (blue), Eq.~\eqref{eqPN:cqfi} (yellow), Eq.~\eqref{eqPN:rqv} (green dashed) and Eq.~\eqref{eqPN:rqfi} (red dashed).
}
\label{fig:SDkpn}
\end{figure}

Since the state is again diagonal in the eigenbasis of $J_z^{\mathrm{B}}$, we obtain
\begin{align}\label{eqPN:rqfi}
\qfi[\rho^{\mathrm{B}},J_z^{\mathrm{B}}]=0.
\end{align}

The data shown in Supplementary Fig.~\ref{fig:SDkpn} shows that the measurement of $S_x^{\mathrm{A}}$ is again optimal for a split twin Fock state, as the conditional Fisher information~(\ref{eqPN:cqfi}) reaches its upper bound~(\ref{eqPN:rqv}).

\section*{Supplementary Note 5 - Bipartite pure states} \label{sec:supp_pure}

\subsection{Witnessing with a fixed $H$}
For a shared pure state we can use the saturation of the inequalities labeled by (*) in~(11), to express the condition~(6) for LHS models as $4\var[\rho^{\mathrm{B}},H] \leq \qfi[\rho^{\mathrm{B}},H]$, whereas in general, this inequality holds in reverse. Hence, steering in this scenario is revealed whenever $4\var[\rho^{\mathrm{B}},H]$ and $\qfi[\rho^{\mathrm{B}},H]$ do not coincide. Even for a fixed $H$ (\ie without optimisation), this condition is close to being a faithful witness of steering: it is satisfied precisely when $H$ is constant on the support of $\rho^{\mathrm{B}}$. This follows from Lemma~\ref{lem:pure_state_vanishing} below.
\begin{lem} \label{lem:pure_state_vanishing}
	$\qfi[\rho,H] = 4 \var[\rho,H]$ if and only if $\Pi_\rho H \Pi_\rho \propto \Pi_\rho$, where $\Pi_\rho$ is the projector on the support of $\rho$.
	\begin{proof}
		Using the spectral decomposition $\rho = \sum_i p_i \proj{i}$ and with $H_{ij}:=\braXket{i}{H}{j}$, we can express~\cite{Toth2018}
		\begin{align} \label{eqn:pure_difference}
			\var[\rho,H]- \frac{1}{4}\qfi[\rho,H] & = 2 \sum_{i\neq j} \frac{p_i p_j}{p_i+p_j} \abs{H_{ij}}^2  + \left[ \sum_i p_i H_{ii}^2 - \left(\sum_i p_i H_{ii} \right)^2 \right],
		\end{align}
		where the second bracketed term, and the terms in the first sum, are all non-negative. When this quantity vanishes, we therefore see that the off-diagonals $H_{ij}=0$ whenever $p_i,p_j \neq 0$. In addition the bracketed term must vanish, and this is simply the variance of the diagonals $H_{ii}$ in the distribution $p_i$. This variance vanishes if and only if the $H_{ii}$ are constant over the range of $i$ such that $p_i \neq 0$. Since
		\begin{align}
			\Pi_\rho & = \sum_{i:\; p_i \neq 0} \proj{i} , \nonumber \\
			\Pi_\rho H \Pi_\rho & = \sum_{\substack{i,j:\\ p_i,p_j\neq 0}} H_{ij} \ketbra{i}{j},
		\end{align}
		these conditions can be equivalently expressed neatly as $\Pi_\rho H \Pi_\rho \propto \Pi_\rho$.
	\end{proof}
\end{lem}

As will be shown in Supplementary Note~\ref{sec:max_avg}, varying over $H$ can make this witness faithful.
When Bob's system is a qubit ($d=2$), without loss of generality we can take the observable to be a Pauli matrix: $H = \mathbf{n}\cdot\boldsymbol{\sigma}$, and then
\begin{equation}
	\cqfi[\psi^{\mathrm{AB}}, \mathbf{n}\cdot\boldsymbol{\sigma}] - 4\cvar[\psi^{\mathrm{AB}}, \mathbf{n}\cdot\boldsymbol{\sigma}] = 8 \left(1 - \tr[(\rho^{\mathrm{B}})^2] \right),
\end{equation}
which is a function of the purity of $\rho^{\mathrm{B}}$ and notably independent of the direction $\mathbf{n}$.

\subsection{Optimal measurements}
In the case of an overall pure state, it is also possible to determine the optimal measurements needed on Alice's side:
\begin{thm}\label{thm:povm}
	i) For a shared pure state $\psi^{\mathrm{AB}}$, an optimal measurement for Alice to achieve $\cqfi[\assem,H]$ is
	\begin{equation}
		\ket{\tilde{x}_k^*} := \frac{1}{\sqrt{d}} \sum_{l=0}^{d-1} e^{2\pi i kl/d} \ket{x_l^*},
	\end{equation}
	where $\ket{x_l}$ are the eigenstates of
	\begin{equation}
		X := \sqrt{\rho^{\mathrm{B}}} H \sqrt{\rho^{\mathrm{B}}} - \expect{H}_{\rho^{\mathrm{B}}} \rho^{\mathrm{B}},
	\end{equation}
	and $*$ denotes complex conjugation in the Schmidt basis of $\psi^{\mathrm{AB}}$.\\

	ii) Similarly, an optimal measurement to achieve $\cvar[\assem,H]$ is $\ket{y_k^*}$, where $\ket{y_k}$ are the eigenstates of the operator
	\begin{equation}
		Y := \sum_i \frac{2\sqrt{p_i p_j}}{p_i + p_j} H_{ij} \ketbra{i}{j}.
	\end{equation}
	\begin{proof}
		We follow the proof of Ref.~\cite{Yu2013}, which found a construction for the optimal pure state ensemble in the concave roof of the variance, but stopped short of giving explicit expressions. As shown there, it is sufficient to find a basis in which the diagonals of $X$ vanish. We show that the basis $\ket{\tilde{x}_k}$ (Fourier transformed with respect to the eigenbasis of $X$) is such a basis. First note that $\tr X = 0$, then writing the spectral decomposition $X = \sum_l x_l \proj{x_l}$,
		\begin{align}
			\braXket{\tilde{x}_k}{X}{\tilde{x}_k} & = \frac{1}{d} \sum_{l,m} e^{2\pi i k (m-l)/d} \braXket{l}{X}{m} \nonumber \\
				& = \frac{1}{d} \sum_l x_l = 0.
		\end{align}
		The remainder proceeds as in Ref.~\cite{Yu2013}, which we include for completeness. The optimal ensemble is constructed by $\sqrt{q_k} \ket{\psi_k} := \sqrt{\rho} \ket{\tilde{x}_k}$ (where we write $\rho^{\mathrm{B}} = \rho$ for brevity). It follows that $q_k = \braXket{\tilde{x}_k}{\rho}{\tilde{x}_k}$, and that this is indeed a valid ensemble decomposition for $\rho$:
		\begin{equation}
			\sum_k q_k \proj{\psi_k} = \sum_k \sqrt{\rho} \proj{\tilde{x}_k} \sqrt{\rho} = \rho.
		\end{equation}
		Now we have
		\begin{align}
			0 = \braXket{\tilde{x}_k}{X}{\tilde{x}_k} & = \bra{\tilde{x}_k} \left( \sqrt{\rho} H \sqrt{\rho} - \expect{H}_\rho \rho \right) \nonumber \\
				& = q_k \braXket{\psi_k}{H}{\psi_k} - \expect{H}_\rho \braXket{\tilde{x}_k}{\rho}{\tilde{x}_k},
		\end{align}
		so $\braXket{\psi_k}{H}{\psi_k} = \expect{H}_\rho$ whenever $q_k \neq 0$. Thus
		\begin{align}
			\sum_k q_k \var[\psi_k,H] & = \sum_k q_k \expect{H^2}_{\psi_k} - \sum_k q_k \expect{H}_{\psi_k}^2 \nonumber \\
				& = \expect{H^2}_{\rho} - \expect{H}_\rho^2 = \var[\rho,H],
		\end{align}
		thus providing the concave roof of the variance.

		The measurement basis for Alice to steer Bob into this ensemble follows straightforwardly. Using the Schmidt decomposition ${\ket{\psi}}_{\mathrm{AB}} = \sum_i \sqrt{p_i} \ket{i} \ket{i}$, the measurement basis for Alice is $\ket{\tilde{x}_k^*}$, where $*$ denotes complex conjugation in the Schmidt basis. This is seen from
		\begin{align}
			{\bra{\tilde{x}_k^*}}_{\mathrm{A}} {\ket{\psi}}_{\mathrm{AB}} & =  \sum_i \sqrt{p_i} \ket{i} {\braket{\tilde{x}_k}{i}}^* \nonumber \\
			& = \sum_i \sqrt{p_i} \ket{i} \braket{i}{\tilde{x}_k} \nonumber \\
			& = \sqrt{\rho} \ket{\tilde{x}_k} \nonumber \\
			& = \sqrt{q_k} \ket{\psi_k}.
		\end{align}

		The corresponding statement for $\cvar$ is similarly given by the optimal convex roof ensemble found in Ref.~\cite{Yu2013}, namely $\sqrt{q_k} \ket{\psi_k} = \sqrt{\rho} \ket{y_k}$. Exactly as above, the measurement required to steer into this ensemble is given by complex conjugation in the Schmidt basis.
	\end{proof}
\end{thm}

\section*{Supplementary Note 6 - Maximal and average violation} \label{sec:max_avg}

Here, we define two quantities involving variation over the generator, namely the maximal and the average violation of the main inequality:

\begin{align}
	\steerMax(\assem) & = \max_{\substack{H \colon \\ \tr[H^2]=1}} \, \pos{ \frac{1}{4}\cqfi[\assem,H] - \cvar[\assem,H] }, \\
    \steerAvg(\assem) & = (d^2-1) \pos{ \int \mu(\dd \mathbf{n}) \frac{1}{4} \cqfi[\assem,\mathbf{n}\cdot\mathbf{H}] - \cvar[\assem,\mathbf{n}\cdot\mathbf{H}]},
\end{align}
where $\pos{\cdot} = \max\{0,\cdot\}$ denotes the positive part, and the $H_i$ provide a basis of $\mathrm{SU}(d)$ generators satisfying $\tr[H_i]=0,\, \tr[H_i H_j] = \delta_{i,j}$, and $\mu$ is the uniform measure over the sphere of unit vectors $\abs{\mathbf{n}}=1$.

Note that both quantities are invariant under unitaries on Bob's side. This is immediately evident for $\steerMax$. To see it for $\steerAvg$, we express the action of some $U$ on the generators as $U^\dagger H_i U = \sum_j R_{i j} H_j$. This action preserves the Hilbert-Schmidt inner product between generators, from which it is found that $R$ must be an orthogonal matrix. Thus $U^\dagger (\mathbf{n}\cdot\mathbf{H}) U = (R^T \mathbf{n})\cdot \mathbf{H}$. Using
\begin{align}
	\cqfi[U^{\mathrm{B}} \assem {U^{\mathrm{B}}}^\dagger, \mathbf{n}\cdot \mathbf{H}] & = \cqfi[\assem, U^\dagger (\mathbf{n}\cdot\mathbf{H}) U] \nonumber \\
		& = \cqfi[\assem, (R^T \mathbf{n})\cdot \mathbf{H}],
\end{align}
and similarly for $\cvar$, it follows that the integral over $\mathbf{n}$ is invariant.\\

We now compute both quantities for a joint pure state $\psi^{\mathrm{AB}}$.

\begin{thm} \label{res:pure_values}
	For a joint pure state $\psi^{\mathrm{AB}}$, we have
	\begin{align} \label{eqn:max_formula}
		\steerMax(\psi^{\mathrm{AB}}) & = \lambda_\mathrm{max}[\mathrm{diag}(\mathbf{p}) - \mathbf{p}\mathbf{p}^T], \\
		\steerAvg(\psi^{\mathrm{AB}}) & =  \sum_{i\neq j} p_i p_j \left( 1 + \frac{2}{p_i+p_j} \right), \label{eqn:avg_formula}
	\end{align}
	where $p_i$ are the eigenvalues of $\rho^{\mathrm{B}}$ and $\lambda_\mathrm{max}[M]$ is the largest eigenvalue of a given matrix $M$.
	\begin{proof}
		From the Methods section, we have, 
		\begin{equation}
		\frac{1}{4} \cqfi[\psi^{\mathrm{AB}},H] - \cvar[\psi^{\mathrm{AB}},H] = \var[\rho^{\mathrm{B}},H] - \frac{1}{4}\qfi[\rho^{\mathrm{B}},H].
		\end{equation}
		This quantity has been studied by T\'oth~\cite{Toth2018}, who shows that
		\begin{align} \label{eqn:v_minus_f}
			\var[\rho,H]- \frac{1}{4}\qfi[\rho,H] & = 2 \sum_{i\neq j} \frac{p_i p_j}{p_i+p_j} \abs{H_{ij}}^2 + \left[ \sum_i p_i H_{ii}^2 - \left(\sum_i p_i H_{ii} \right)^2 \right],
		\end{align}
		and computes the average needed for $\steerAvg$.
		
		For $\steerMax$, we turn \eqref{eqn:v_minus_f} into a matrix expression. We encode the components $H_{ij}$ into a vector whose first $d$ components are the diagonals and remaining $d(d-1)/2$ components are the off-diagonals:
		\begin{equation}
			\mathbf{v} = (H_{11},H_{22},\dots, \sqrt{2} H_{12}, \sqrt{2} H_{13},\dots)
		\end{equation}
		such that $\abs{\mathbf{v}}^2 = \sum_{ij} \abs{H_{ij}}^2 = \tr[H^2] = 1$. Similarly define a matrix $M = M_D \oplus M_O$ split into diagonal and off-diagonal parts:
		\begin{align}
			[M_D]_{i,j} & = \delta_{i,j} p_i - p_i p_j, \\
			[M_O]_{(ij),(kl)} & = \delta_{i,k} \delta_{j,l} \left( \frac{2p_i p_j}{p_i+p_j} \right).
		\end{align}
		We then see that
		\begin{align}
			\var[\rho,H]-\frac{1}{4}\qfi[\rho,H] & = \mathbf{v}^\dagger M \mathbf{v} \nonumber	 \\
			\Rightarrow \steerMax(\psi^{\mathrm{AB}}) & = \lambda_\mathrm{max}[M].
		\end{align}
		Finally, we will show that $\lambda_\mathrm{max}[M] = \lambda_\mathrm{max}[M_D] \geq \lambda_\mathrm{max}[M_O]$. Note that the diagonals of $M_O$ are the harmonic means of each pair of $p_i,p_j,\, (i\neq j)$, and so
		\begin{equation} \label{eqn:off-diag_eigenval}
			\lambda_\mathrm{max}[M_O] = \frac{2p_1 p_2}{p_1 + p_2} =: m_*,
		\end{equation}
		where without loss of generality we have ordered $p_1 \geq p_2 \geq \dots \geq p_d$. Note that $m_* \geq p_2$.
		
		From the Weyl inequalities on eigenvalues \cite[Theorem III.2.1]{Bhatia1997}, we can bound
		\begin{align}
			\lambda_\mathrm{max}[M_D] & \leq \lambda_\mathrm{max}[\mathrm{diag}(\mathbf{p})] + \lambda_\mathrm{max}[-\mathbf{p}\mathbf{p}^T] \nonumber \\
				& = p_1, \\
			\lambda_\mathrm{max}[M_D] & \geq \lambda_2[\mathrm{diag}(\mathbf{p})] + \lambda_{d-1}[-\mathbf{p}\mathbf{p}^T] \nonumber \\
				& = p_2,
		\end{align}
		where $\lambda_k$ denotes the $k$th largest eigenvalue. So we have $p_2 \leq \lambda_\mathrm{max}[M_D] \leq p_1$ -- we will now obtain a stronger lower bound. Let us inspect the eigenvector condition $M_D \mathbf{w} = m \mathbf{w}$:
		\begin{align} \label{eqn:eigenvector}
			p_i w_i - p_i \sum_j p_j w_j & = m w_i \quad \forall i \nonumber \\
			\Rightarrow w_i = \frac{p_i \bar{w}}{p_i-m}, \; \bar{w} = \sum_j p_j w_j.
		\end{align}
		This determines the eigenvector corresponding to the eigenvalue $m$. Multiplying by $p_i$ and summing over $i$,
		\begin{align}
			\bar{w} & = \sum_i \frac{p_i^2 \bar{w}}{p_i-m} \nonumber \\
			\Rightarrow \bar{w} & = 0 \text{ or } \sum_i \frac{p_i^2}{p_i-m} = 1.
		\end{align}
		Now $\bar{w} \neq 0$ or else we would have $\mathbf{w}=\mathbf{0}$ by \eqref{eqn:eigenvector}. Therefore $m$ satisfies $g(m) := \sum_i p_i^2/(p_i-m) = 1$. The function $g$ is strictly increasing in regions where it is continuous:
		\begin{equation}
			g'(m) = \sum_i \frac{p_i^2}{(p_i-m)^2} > 0.
		\end{equation}
		Evaluating $g(m_*)$ from \eqref{eqn:off-diag_eigenval},
		\begin{align}
			g(m_*) & = \frac{p_1^2}{p_1-2p_1 p_2/(p_1+p_2)} + \frac{p_2^2}{p_2-2p_1 p_2/(p_1+p_2)} \nonumber \\
				& \quad + \sum_{i>2} \frac{p_i^2}{p_i-m_*} \nonumber \\
				& = (p_1+p_2) \left( \frac{p_1}{p_1-p_2} + \frac{p_2}{p_2-p_1} \right) + \sum_{i>2} \frac{p_i^2}{p_i-m_*} \nonumber \\
				& = p_1 + p_2 + \sum_{i>2} \frac{p_i^2}{p_i-m_*} \nonumber \\
				& \leq p_1 + p_2 \nonumber \\
				& \leq 1,
		\end{align}
		since $m_* \geq p_2 \geq p_3 \geq \dots$. We are working in the region $p_2 \leq m \leq p_1$ as shown above, in which $g$ is continuous and thus strictly increasing. So in order to have $g(m)=1$, it must be that $m\geq m_*$. In other words, $\lambda_\mathrm{max}[M_D] \geq \lambda_\mathrm{max}[M_O]$.
	\end{proof}
\end{thm}

\begin{figure}[h]
	\includegraphics[width=.6\textwidth]{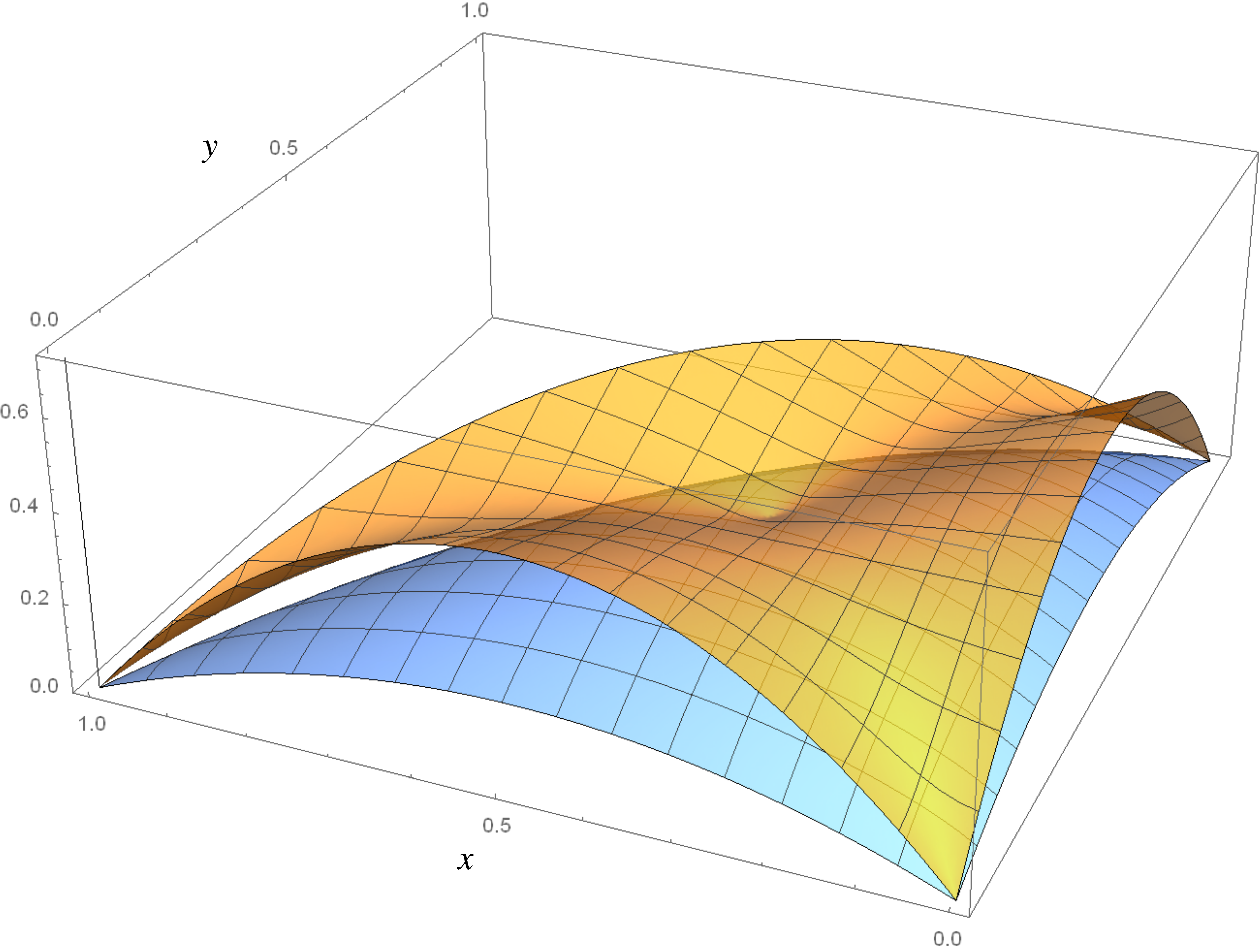}
	\caption{\textbf{Steering quantifiers for pure bipartite states.} A plot of the quantities $\steerAvg/(d^2-1)$ (blue) and $\steerMax$ (orange) for a pure state with Schmidt coefficients $x,\, y,\, 1-x-y$. Apart from the extreme points, using this normalisation they coincide at $x=y=1/3$, corresponding to the maximally entangled state with $d=3$.}
	\label{fig:measures}
\end{figure}

For pure states, steering is equivalent to entanglement, therefore any steering measure must reduce to an entanglement measure when evaluated on pure states~\cite{WisemanPRL2007,GisinPLA1991}.

\begin{cor}
	$\steerAvg$ is a full entanglement measure for pure states. $\steerMax$ is a faithful witness of entanglement for pure states, but not a monotone under LOCC.
	\begin{proof}
		It is easy to see that $\steerAvg$ vanishes if and only if $p_i p_j=0 \; \forall i \neq j$, \ie $p_1=1,\,p_2=p_3=\dots=0$, meaning that $\rho^{\mathrm{B}}$ is pure.
		
        A function $f(\rho^{\mathrm{B}})$ satisfies strong monotonicity under LOCC if and only if (i) $f$ is a symmetric function of the eigenvalues of $\rho^{\mathrm{B}}$, (ii) $f$ is expansible (meaning that zero eigenvalues can be appended without changing the value) and (iii) $f$ is concave~\cite{Vidal2000,Horodecki2009}. Properties (i) and (ii) are evident from the expression \eqref{eqn:avg_formula}, while (iii) follows from concavity of $V$ and convexity of $\qfi$.\\
        
        $\steerMax$ vanishes if and only if $\psi^{\mathrm{AB}}$ is separable, and it is a symmetric, expansible function of $\mathbf{p}$. However, it is not an entanglement monotone, since its maximal value is attained for $\mathbf{p}=(1/2,1/2,0,\dots)$ (observed numerically; see Supplementary Fig.~\ref{fig:measures}), which in $d>2$ does not correspond to the maximally entangled state.
	\end{proof}
\end{cor}

For general mixed states, the values in Theorem~\ref{res:pure_values} are upper bounds to $\steerAvg,\, \steerMax$.\\

Now we characterise further aspects of these quantities applied to general assemblages. At this stage, it is convenient to introduce the notation
\begin{equation}
    \Delta^{\mathrm{B}|\mathrm{A}}[\assem,H] := \frac{1}{4}\cqfi[\assem,H]-\cvar[\assem,H].
\end{equation}

\begin{lem} \label{lem:convexity}
    $\steerMax$ and $\steerAvg$ are convex.
    \begin{proof}
        We first show that $\cqfi[\assem,H]$ is convex in $\assem$. By definition, a convex combination of two assemblages takes the form $[q\assem + (1-q)\assem'](a|X) = q \assem(a|X) + (1-q)\assem'(a|X) = q p(a|X)\rho_{a|X} + (1-q) p'(a|X)\rho'_{a|X}$. From convexity of the QFI, we see
        \begin{align}
            \cqfi[q\assem + (1-q)\assem',H] & \leq \max_X\left( \sum_a q p(a|X) \qfi[\rho_{a|X},H] + (1-q)p'(a|X) \qfi[\rho'_{a|X},H]\right) \nonumber \\
                & \leq q \max_X \sum_a p(a|X) \qfi[\rho_{a|X},H] + (1-q) \max_{X} \sum_a p'(a|X) \qfi[\rho'_{a|X},H] \nonumber \\
                & = q \cqfi[\assem,H] + (1-q) \cqfi[\assem',H].
        \end{align}
        The same argument shows that $\cvar$ is concave in $\assem$, and so $\Delta^{\mathrm{B}|\mathrm{A}}$ is also convex.\\
        
        For $\steerMax$, we also need to employ convexity of the function $\pos{\cdot}$:
        \begin{align}
            \steerMax(q\assem + [1-q]\assem') & \leq \max_H \pos{ q\Delta^{\mathrm{B}|\mathrm{A}}[\assem,H] + (1-q)\Delta^{\mathrm{B}|\mathrm{A}}[\assem',H] } \nonumber \\
                & \leq \max_H \left(q \pos{ \Delta^{\mathrm{B}|\mathrm{A}}[\assem,H] } + (1-q) \pos{ \Delta^{\mathrm{B}|\mathrm{A}}[\assem',H]} \right)\nonumber \\
                & \leq q \max_H \Delta^{\mathrm{B}|\mathrm{A}}[\assem,H] + (1-q) \max_{H} \Delta^{\mathrm{B}|\mathrm{A}}[\assem',H] \nonumber \\
                & = q \steerMax(\assem) + (1-q) \steerMax(\assem').
        \end{align}
        The reasoning is very similar for $\steerAvg$:
        \begin{align}
            \steerAvg(q\assem+[1-q]\assem') & \leq (d^2-1) \pos{ \int \mu(\dd \mathbf{n}) \, q \Delta^{\mathrm{B}|\mathrm{A}}[\assem,\mathbf{n}\cdot\mathbf{H}] + (1-q)\Delta^{\mathrm{B}|\mathrm{A}}[\assem',\mathbf{n}\cdot\mathbf{H}] } \nonumber \\
                & \leq q (d^2-1)\pos{ \int \mu(\dd \mathbf{n})\, \Delta^{\mathrm{B}|\mathrm{A}}[\assem,\mathbf{n}\cdot\mathbf{H}] } + (1-q)(d^2-1) \pos{ \int \mu(\dd \mathbf{n})\, \Delta^{\mathrm{B}|\mathrm{A}}[\assem',\mathbf{n}\cdot\mathbf{H}] } \nonumber \\
                & = q \steerAvg(\assem) + (1-q) \steerAvg(\assem').
        \end{align}
    \end{proof}
\end{lem}

The following results demonstrate an aspect of these quantities that is well behaved: being unchanged whenever Bob appends an additional pure state $\phi = \proj{\phi}$ to his side. This is a necessary condition for monotonicity under 1W-LOCC, and may be a useful step in proving that, if true.

\begin{lem} \label{lem:ancilla}
	Suppose an ancillary system B' in a pure state $\phi$ is appended to Bob's side -- the new assemblage is denoted $\assem \ox \phi^{\mathrm{B}}$. Then, for any observable $H$ acting on the system BB$'$, we have
    \begin{equation}
        \Delta^{\mathrm{BB}'|\mathrm{A}}[\assem \ox \phi^{\mathrm{B}'}, H] = \Delta^{\mathrm{B}|\mathrm{A}}[\assem, \tilde{H}],
    \end{equation}
	where $\tilde{H} := {\bra{\phi}}_{B'} H {\ket{\phi}}_{B'}$.
	\begin{proof}
		Lemma~3 from Ref.~\cite{Morris2020} says that, for any projector $\Pi$ with the property $\Pi \rho = \rho$,
		\begin{align}
			\qfi[\rho,H] - 4\var[\rho,H] = \qfi[\rho,\Pi H \Pi] - 4\var[\rho,\Pi H \Pi].
		\end{align}
		We take $\rho = \rho_{a|X}\ox\phi^{\mathrm{B}'},\, \Pi = I \ox \phi^{\mathrm{B}'}$, such that $\Pi H \Pi = \tilde{H} \ox \phi^{\mathrm{B}'}$, and then 
		\begin{align}
			\qfi[\rho_{a|X}\ox \phi^{\mathrm{B}'},H] & = \qfi[\rho_{a|X}\ox \phi^{\mathrm{B}'},\tilde{H} \ox \phi^{\mathrm{B}'}] + 4\var[\rho_{a|X} \ox \phi^{\mathrm{B}'}, H] - 4\var[\rho_{a|X} \ox \phi^{\mathrm{B}'}, \tilde{H} \ox \phi^{\mathrm{B}'}] \nonumber \\
				& = \qfi[\rho_{a|X},\tilde{H}] + 4\var[\rho_{a|X} \ox \phi^{\mathrm{B}'},H] - 4\var[\rho_{a|X}, \tilde{H}] \nonumber \\
				& = \qfi[\rho_{a|X},\tilde{H}] + 4\expect{H^2}_{\rho_{a|X}\ox\phi^{\mathrm{B}'}} - 4\expect{H}_{\rho_{a|X}\ox\phi^{\mathrm{B}'}}^2 - 4\expect{\tilde{H}^2}_{\rho_{a|X}} + 4\expect{\tilde{H}}_{\rho_{a|X}}^2 \nonumber \\
				& = \qfi[\rho_{a|X},\tilde{H}] + 4 \expect{{\bra{\phi}}_{B'}H^2{\ket{\phi}}_{B'} - \tilde{H}^2}_{\rho_{a|X}},
		\end{align}
		resulting in
		\begin{align}
			\cqfi[\assem\ox \phi^{\mathrm{B}'},H] & = \max_X \, \sum_a p(a|X) \left( \qfi[\rho_{a|X},\tilde{H}] + 4\expect{{\bra{\phi}}_{B'}H^2{\ket{\phi}}_{B'} - \tilde{H}^2}_{\rho_{a|X}} \right) \nonumber \\
				& = \max_X \, \sum_a p(a|X) \qfi[\rho_{a|X},\tilde{H}] + 4\expect{{\bra{\phi}}_{B'}H^2{\ket{\phi}}_{B'} - \tilde{H}^2}_{\rho^{\mathrm{B}}} \nonumber \\
				& = \cqfi[\assem,\tilde{H}] + 4 \expect{{\bra{\phi}}_{B'}H^2{\ket{\phi}}_{B'} - \tilde{H}^2}_{\rho^{\mathrm{B}}}.
		\end{align}
		Similarly,
		\begin{align}
			\cvar[\assem\ox\phi^{\mathrm{B}'},H] & = \min_X \, \sum_a p(a|X) \var[\rho_{a|X}\ox\phi^{\mathrm{B}'}, H] \nonumber \\
				& = \min_X \, \sum_a p(a|X) \left( \expect{ {\bra{\phi}}_{B'}H^2{\ket{\phi}}_{B'}}_{\rho_{a|X}} - \expect{\tilde{H}}_{\rho_{a|X}}^2 \right) \nonumber \\
				& = \min_X \, \sum_a p(a|X) \left( \expect{ {\bra{\phi}}_{B'}H^2{\ket{\phi}}_{B'} - \tilde{H}^2}_{\rho_{a|X}} + \var[\rho_{a|X},\tilde{H}] \right) \nonumber \\
				& = \expect{ {\bra{\phi}}_{B'}H^2{\ket{\phi}}_{B'} - \tilde{H}^2}_{\rho^{\mathrm{B}}} + \min_X \, \sum_a p(a|X) \var[\rho_{a|X},\tilde{H}] \nonumber \\
				& = \expect{ {\bra{\phi}}_{B'}H^2{\ket{\phi}}_{B'} - \tilde{H}^2}_{\rho^{\mathrm{B}}} + \cvar[\assem,\tilde{H}].
		\end{align}
		Putting these last two equations together gives the claimed result.
	\end{proof}
\end{lem}

\begin{lem}
	$\steerMax(\assem\ox\phi^{\mathrm{B}'}) = \steerMax(\assem)$
	\begin{proof}
		Take any $H$ on BB$'$ such that $\tr[H]=0,\, \tr[H^2]=1$. Picking any product basis such that ${\ket{0}}_{B'} = {\ket{\phi}}_{B'}$, we have
		\begin{align}
			\tr[H^2] & = \sum_{i,j,k,l}  \abs{\braXket{ik}{H}{jl}}^2 \nonumber \\
				& \geq \sum_{i,j} \abs{\braXket{i0}{H}{j0}}^2 \nonumber \\
				& = \sum_{i,j} \abs{\braXket{i}{\tilde{H}}{j}}^2 \nonumber \\
				& = \tr[\tilde{H}^2].
		\end{align}
		Therefore $\tr[\tilde{H}^2] \leq 1$; however, we need not have $\tr[\tilde{H}]=0$. By appropriately shifting and scaling, we have a new generator
		\begin{equation}
			G := \frac{\tilde{H} - \tr[\tilde{H}]I }{\tr[\tilde{H}^2]}
		\end{equation}
		satisfying $\tr[G]=0,\, \tr[G^2]=1$. (This is all assuming $\tilde{H} \neq 0$, otherwise the remainder of the argument is trivial.) From Lemma~\ref{lem:ancilla}, picking the optimal $H$ for witnessing steering on $ABB'$,
		\begin{align}
			\steerMax(\assem\ox\phi^{\mathrm{B}'}) & = \frac{1}{4}\qfi^{\mathrm{BB}'|\mathrm{A}}[\assem\ox\phi^{\mathrm{B}'},H] - \var^{\mathrm{BB}'|\mathrm{A}}[\assem\ox\phi^{\mathrm{B}'},H] \nonumber \\
				& = \frac{1}{4}\cqfi[\assem,\tilde{H}] - \cvar[\assem,\tilde{H}] \nonumber \\
				& = \tr[\tilde{H}^2] \left( \frac{1}{4}\cqfi[\assem,G] - \cvar[\assem,G] \right) \nonumber \\
				& \leq \frac{1}{4}\cqfi[\assem,G] - \cvar[\assem,G] \nonumber \\
				& \leq \steerMax(\assem).
		\end{align}
		The reverse inequality is easily seen by noting that the set of generators on BB$'$ includes those of the form $H^{\mathrm{B}} \ox I^{\mathrm{B}'}$, and so 
		\begin{align}
			\steerMax(\assem\ox\phi^{\mathrm{B}'}) & \geq \max_{H^{\mathrm{B}}} \left(\frac{1}{4}\qfi^{\mathrm{BB}'|\mathrm{A}}[\assem\ox\phi^{\mathrm{B}'},H^{\mathrm{B}}\ox I^{\mathrm{B}'}] - \var^{\mathrm{BB}'|\mathrm{A}}[\assem\ox\phi^{\mathrm{B}'},H^{\mathrm{B}} \ox I^{\mathrm{B}'}]\right) \nonumber \\
				& = \max_{H^{\mathrm{B}}} \left(\frac{1}{4}\cqfi[\assem,H^{\mathrm{B}}] - \cvar[\assem,H^{\mathrm{B}}]\right) \nonumber \\
				& = \steerMax(\assem).
		\end{align}
	\end{proof}
\end{lem}

\begin{lem}
	$\steerAvg(\assem \ox \phi^{\mathrm{B}'}) = \steerAvg(\assem)$
	\begin{proof}

        
        Let B and B$'$ have respective dimensions $d$ and $d'$. B thus has a set $\bg{h}$ of orthonormal $\mathrm{SU}(d)$ generators $h_\mu,\, \mu=1,\dots,g:= d^2-1$, and similarly B$'$ has $g':={d'}^2-1$ generators $\bg{h}'$. It is convenient to set $h_0 = I/\sqrt{d}$, which completes the orthonormal basis of Hermitian generators. For the joint system BB$'$, we can choose a set $\bg{H}$ of $G:= (dd')^2-1$ generators $H_{\mu,\nu} := h_\mu \ox h'_\nu$ where $\mu = 0,1,\dots,g,\, \nu=0,1,\dots,g'$, but $\mu=\nu=0$ is excluded (since we are only interested in operators with zero trace).

		Applying Lemma~\ref{lem:ancilla}, we have $\Delta^{\mathrm{BB}'|\mathrm{A}}[\assem \ox \phi^{\mathrm{B}'}, \bg{N}\cdot\bg{H}] = \Delta^{\mathrm{B}|\mathrm{A}}[\assem,\bg{N}\cdot\tilde{\bg{H}}]$, where $\tilde{H}_{\mu,\nu} = \braXket{\phi}{h_\nu}{\phi} h_\mu$. Then
		\begin{align}
			\bg{N}\cdot\bg{\tilde{H}} & = \sum_{\mu=0}^g \sum_{\nu=0}^{g'} N_{\mu,\nu} \braXket{\phi}{h_\nu}{\phi} h_\mu \nonumber \\
				& = \sum_{\mu=1}^g \sum_{\nu=1}^{g'} N_{\mu,\nu} \braXket{\phi}{h_\nu}{\phi}h_\mu + \sum_{\mu=1}^g \frac{N_{\mu,0}}{\sqrt{d'}} h_\mu + \sum_{\nu=1}^{g'} \frac{\braXket{\phi}{h_\nu}{\phi}}{\sqrt{d}} I.
		\end{align}
		The constant term does not contribute to the QFI or the variance, so 
		\begin{align}
			\Delta^{\mathrm{BB}'|\mathrm{A}}[\assem \ox \phi^{\mathrm{B}'}, \bg{N}\cdot\bg{H}] & = \Delta^{\mathrm{B}|\mathrm{A}}[\assem, \bg{n} \cdot \bg{h}], \nonumber \\
				n_\mu & = \sum_{\nu=0}^{g'} \braXket{\phi}{h_\nu}{\phi} N_{\mu,\nu}.
		\end{align}

		In order to perform the integral, we rotate to a convenient choice of coordinates $M_{\mu\nu} = R_{\mu\nu,\lambda\,\sigma} N_{\lambda\sigma}$. $R$ is chosen to be an orthogonal matrix with the components in the first $g$ rows set to $R_{\mu 0, \lambda\sigma} := \delta_{\mu,\lambda} \braXket{\phi}{h_\sigma}{\phi}$. This is possible because of the orthonormality of the vectors $\bg{v}^\mu:= (\delta_{\mu,\lambda} \braXket{\phi}{h_\sigma}{\phi})_{\lambda,\sigma}$ (it is easy to show that $\sum_\sigma {\braXket{\phi}{h_\sigma}{\phi}}^2=1$). These coordinates are such that $M_{\mu 0} = n_\mu$.

		Letting $A(D)$ denote the surface area of a $D$-dimensional sphere, we have
		\begin{align}
			\steerAvg(\assem\ox\phi^{\mathrm{B}'}) & = \frac{G}{A(G-1)} \pos{\int_{\abs{\bg{N}}=1} \dd \bg{N} \, \Delta^{\mathrm{BB}'|\mathrm{A}}[\assem\ox\phi^{\mathrm{B}'},\bg{N}\cdot\bg{H}] }\nonumber \\
				& = \frac{G}{A(G-1)} \pos{ \int_{\abs{\bg{M}}=1} \dd \bg{M} \, \Delta^{\mathrm{B}|\mathrm{A}}[\assem, \sum_\mu M_{\mu 0}h_\mu ] } .
		\end{align}
		Now we employ generalised spherical coordinates, where $G$ cartesian coordinates $x_i,\, i=1,\dots,G$ are represented in terms of angles $\theta_1,\dots,\theta_{G-1}$ (and a radius which here is $r=1$):
		\begin{align}
			x_1 & = \cos \theta_1, \nonumber \\
			x_2 & = \sin \theta_1 \cos \theta_2, \nonumber \\
			\vdots & \nonumber \\
			x_{G-1} & = \sin \theta_1 \sin \theta_2 \dots \sin \theta_{G-2} \cos \theta_{G-1}, \nonumber \\
			x_G & = \sin \theta_1 \sin \theta_2 \dots \sin \theta_{G-2} \sin \theta_{G-1},
		\end{align}
		where all $\theta_i$ are in the range $[0,\pi)$ apart from $\theta_{G-1} \in [0,2\pi)$. The volume element is $\dd \bg{x} = \prod_{i=1}^G \sin^{G-i-1}(\theta_i) \dd \theta_i$. We set the $M_{\mu 0}$ to equal the last $g$ of these coordinates, i.e., $M_{\mu 0} = x_{G-g+\mu}$.
		
		It is easy to show that $\sum_{\mu=1}^g M_{\mu 0}^2 = \sum_{\mu=1}^g x_{G-g+\mu}^2 = \prod_{i=1}^{G-g} \sin^2 \theta_i$. Thus we can normalise $\sum_\mu M_{\mu 0}h_\mu$ by dividing by this norm, obtaining
		\begin{align}
			\hat{\bg{n}}\cdot \bg{h} & := \frac{1}{\sqrt{\sum_\mu M_{\mu 0}^2}} \sum_\mu M_{\mu 0}h_\mu \nonumber \\
			& = \pm \cos (\theta_{G-g+1}) h_1 \pm \sin (\theta_{G-g+1}) \cos(\theta_{G-g+2}) h_2 + \dots \pm \sin(\theta_{G-g+1})\dots\sin(\theta_{G-1}) h_g.
		\end{align}
		From this, we see that $\hat{\bg{n}}$ varies over all unit vectors in the subspace spanned by the final $g$ cartesian coordinates. By factoring the norm outside the QFI and variance, we have
		\begin{align}
			\steerAvg(\assem\ox \phi^{\mathrm{B}'}) & = \frac{G}{A(G-1)} \pos{ \int \left( \prod_{i=1}^G \sin^{G-i-1}(\theta_i) \dd \theta_i \right) \left(\prod_{i=1}^{G-g} \sin^2(\theta_i) \right) \Delta^{\mathrm{B}|\mathrm{A}}[\assem,\hat{\bg{n}}\cdot \bg{h}] } \nonumber \\
				& = \frac{G}{A(G-1)} \pos{ \int \left( \prod_{i=1}^{G-g} \sin^{G-i+1}(\theta_i) \dd \theta_i \right) \int \left( \prod_{i=G-g+1}^G \sin^{G-i-1}(\theta_i) \dd \theta_i \right) \Delta^{\mathrm{B}|\mathrm{A}}[\assem,\hat{\bg{n}}\cdot \bg{h}] } \nonumber \\
				& = \frac{G}{A(G-1)} \pos{ \left( \prod_{i=1}^{G-g} \int \sin^{G-i+1}(\theta) \dd \theta \right) \frac{A(g-1)}{g} \steerAvg(\assem) }.
		\end{align}
		Now
		\begin{align}
			\int \sin^k (\theta) \dd \theta & = \frac{\pi^{1/2} \Gamma\left(\frac{k+1}{2}\right)}{\Gamma\left(\frac{k+2}{2}\right)}, \nonumber \\
			A(g-1) & = \frac{2\pi^{g/2}}{\Gamma\left(\frac{g}{2}\right)}, 
		\end{align}
		so that
		\begin{align}
			\steerAvg(\assem\ox\phi^{\mathrm{B}'}) & = \frac{G A(g-1)}{g A(G-1)} \pi^{(G-g)/2} \frac{\Gamma\left( \frac{G+1}{2} \right)}{\Gamma\left( \frac{G+2}{2} \right)} \cdot \frac{\Gamma\left( \frac{G}{2} \right)}{\Gamma\left( \frac{G+1}{2} \right)} \dots \frac{\Gamma\left( \frac{g+4}{2} \right)}{\Gamma\left( \frac{g+3}{2} \right)} \cdot \frac{\Gamma\left( \frac{g+2}{2} \right)}{\Gamma\left( \frac{g+3}{2} \right)} \steerAvg(\assem) \nonumber \\
				& = \frac{2G \pi^{g/2} \Gamma\left( \frac{G}{2} \right) }{2g \pi^{G/2} \Gamma\left( \frac{g}{2}\right)} \pi^{(G-g)/2} \cdot \frac{\Gamma\left( \frac{g}{2}+1\right)}{\Gamma\left( \frac{G}{2}+1\right)} \steerAvg(\assem) \nonumber \\
				& = \frac{\frac{G}{2} \Gamma\left( \frac{G}{2}\right)}{\frac{g}{2} \Gamma\left( \frac{g}{2}\right)} \cdot \frac{\Gamma\left( \frac{g}{2}+1\right)}{\Gamma\left( \frac{G}{2}+1\right)} \steerAvg(\assem) \nonumber \\
				& = \steerAvg(\assem),
		\end{align}
		having used the identity $x \Gamma(x) = \Gamma(x+1)$.
	\end{proof}
\end{lem}

\end{document}